\documentclass[acmlarge]{acmart}

\makeatletter
\newcommand{\confshort}{\acmConference@shortname}
\newcommand{\conffull}{\acmConference@name}
\newcommand{\confdate}{\acmConference@date}
\newcommand{\confloc}{\acmConference@venue}
\AtBeginDocument{%
  \fancypagestyle{firstpagestyle}{%
    \fancyhead{}%
    \fancyfoot[C]{\@headfootfont\footnotesize\thepage}%
  }%
  \fancyhf{}%
  \fancyhead[LO]{\@headfootfont\shorttitle}%
  \fancyhead[RE]{\@headfootfont\@shortauthors}%
  \fancyhead[LE]{\@headfootfont\footnotesize \confshort, \confdate, \confloc}%
  \fancyhead[RO]{\@headfootfont\footnotesize \confshort, \confdate, \confloc}%
  \fancyfoot[C]{\@headfootfont\footnotesize\thepage}%
}
\makeatother
\acmBooktitle{\conffull\@ (\confshort), \confdate, \confloc}

\AtBeginDocument{%
  }

\usepackage{enumitem}
\usepackage[textsize=footnotesize]{todonotes}
\usepackage{cleveref}
\usepackage{tabulary}
\usepackage{longtable}
\usepackage{booktabs}

\usepackage{amssymb}
\usepackage{pifont}
\newcommand{\cmark}{\ding{51}}
\newcommand{\xmark}{\ding{55}}
\usepackage{array}
\usepackage{tabularx}

\setcopyright{none}
\makeatletter
\long\def\footnotetextcopyrightpermission#1{}
\makeatother

\acmConference[FAccT '26]{The 2026 ACM Conference on Fairness, Accountability, and Transparency}{June 25--28, 2026}{Montreal, QC, Canada}
\acmBooktitle{The 2026 ACM Conference on Fairness, Accountability, and Transparency (FAccT '26), June 25--28, 2026, Montreal, QC, Canada}

\begin{document}

\title{Barriers to Evidence in AI-Related Cases and the Privatization of Proof}

\author{Sarah H. Cen}
\authornote{All authors contributed equally to this research.}
\email{sarahcen@andrew.cmu.edu}
\affiliation{%
  \institution{Carnegie Mellon University}
  \city{Pittsburgh}
  \country{USA}
}

\author{Hannah Ismael}
\email{hannahismael42@berkeley.edu}
\authornotemark[1]
\affiliation{%
  \institution{University of California, Berkeley}
  \city{Berkeley}
  \country{USA}
}

\author{Lucia Zheng}
\authornotemark[1]
\affiliation{%
  \institution{Stanford University}
  \city{Palo Alto}
  \country{USA}}
\email{zlucia@stanford.edu}

\renewcommand{\shortauthors}{Cen, Ismael, and Zheng}

\makeatletter
\def\@authorsaddresses{%
  \@mkauthorsaddresses
  \endgraf
  \vskip 0.75\baselineskip
  \noindent
  To appear in the Proceedings of the 2026 ACM Conference on Fairness, Accountability, and Transparency (FAccT '26), June 25--28, 2026, Montreal, QC, Canada.}
\makeatother

\begin{abstract}
  Evidence lies at the core of litigation, but it is increasingly difficult to obtain in AI-related disputes. Even when a claimant's position has merit, cases are often settled or dismissed because decisive facts are hidden inside proprietary models, platform logs, and protected databases. Grounding our discussion in past and ongoing cases, we investigate how asymmetries in access, resources, and expertise can create significant barriers to evidence in AI-related cases. We show how developers and deployers resist disclosure through various strategies challenging the value of the evidence to the requesting party and the cost of evidence production. From these patterns we identify seven recurring sources of asymmetry---access to models, data, documentation, logs, expertise, compute, and infrastructure---that reflect a broader pattern that we call the privatization of proof: when control over proof falls in the hands of private actors that can demand justification for access while ensuring that justification remains out of reach.
  We further argue that different types of access can be fungible: in the absence of a certain type of access (e.g., to model internals), one may be able to use alternative forms of access (e.g., sufficient compute, query access, and access to user logs) and to obtain a functionally equivalent amount of information.
  We propose a three-part test that can help resolve AI access disputes in litigation, drawing on concepts such as proportionality and reasonable alternatives.
  Our test relies on a few observations, including that the cause of action can provide a baseline for access.
\end{abstract}


\received{17 May 2026}
\received[revised]{25 March 2026}
\received[accepted]{16 April 2026}

\maketitle

\section{Introduction}\label{sec:intro}

Evidence lies at the core of every legal system. 
Without the ability to prove or defend against a claim, 
legal action cannot be meaningfully pursued.
Yet evidence is difficult to acquire. 
For a claim to be true is not enough.
Supporting information must be observed and captured, often even measured and quantified. 
Because legal decisions hinge on key evidence, parties who may be harmed by it guard it closely. 
They may do so not only by limiting ways to obtain information, but also by restricting the conditions under which information that is already in hand can be turned into usable proof (e.g., substantiated and validated).

The treatment of evidence is governed by a complex system of doctrines.
Among these is the burden of proof, which assigns responsibility for raising issues, producing evidence, and ultimately persuading a factfinder (e.g., judge or jury). While the burden of proof (and who bears it) plays an important role in determining outcomes, it does not capture the full picture. 
In practice, evidence is produced and contested in a repeated back-and-forth throughout legal proceedings. 
During this process, procedural rules determine whether parties can obtain access to relevant information, 
while evidentiary rules and gatekeeping standards determine whether that information can be developed into evidence. 
Together, these doctrines 
shape how claims are supported and contested.

Parties challenging AI-related systems and actions often face barriers to evidence that can be insurmountable in practice because the access, expertise, and resources needed to satisfy them are unevenly distributed. 
This dynamic should be concerning even to those who do not directly use AI. 
It implies that AI-related decisions and outcomes may become uncontestable without special access, support, or knowledge.
Yet private litigation remains a critical route to curbing AI risks, especially in the absence of government action.

In AI-related disputes, both obtaining relevant information and turning it into admissible, reliable, and ultimately persuasive evidence are increasingly difficult.

On the production side, access to AI is increasingly restricted due to claims of trade secrecy, contractual obligations, and user privacy.
Even if access is granted, the production of useful evidence typically requires significant resources and expert knowledge, 
but both are increasingly concentrated within AI companies or costly to obtain.
There are also issues of preservation: for instance, an AI model may be updated or deprecated, preventing an examination of the specific model responsible for the behavior in question. 

On the validation side, even when the challenging party already has relevant information, 
they must substantiate it with evidence that courts treat as reliable and admissible.
Ultimately, this evidence must be compelling enough to persuade the factfinder.
Yet both validation and persuasion run into similar barriers: restricted access to key information, 
limited resources to test and reproduce AI behavior, and increasingly few independent experts. 
These difficulties are complicated when the AI developer/controller is a nonparty (i.e., neither the plaintiff nor the defendant) and therefore subject to 
stronger protections against compelled access, which occurs when AI is procured rather than developed in-house. 

These difficulties are not unique to AI. Courts have long confronted asymmetries in access, especially to complex technical evidence. 
Disagreements over trade secrecy, confidentiality, cost and burden of discovery, and preservation are common. 
As similar barriers arise in AI-related cases (such as the ongoing cases over \emph{Mobley v. Workday} and \emph{Lokken v. UnitedHealth Group}), 
it becomes important to systematically investigate how AI reshapes dynamics around evidence to avoid a future in which AI oversight hinges on special access, support, or knowledge. 
In this Article, we investigate when and why AI-related evidence may be difficult to obtain and support.

\paragraph{Contributions.} 
Across the cases we examine, we find that barriers to evidence frequently determine whether claims challenging an AI system or action can proceed at all, with many being resolved or settled in the pre-trial stage.  
Asymmetries in information, access, resources, and expertise surface early in litigation, when claimants must make preliminary showings to obtain discovery.
When such showings cannot be made, cases often stall.
When courts have compelled disclosure, clear flaws in AI system behavior or design have often led to settlements.

In \Cref{sec:strategies}, 
we describe arguments the producing party uses to resist the requesting party's efforts to obtain evidence that is within producing party's control in cases concerning AI and other automated technologies. 
We observe two recurring categories of arguments: (1) that the requesting party has not sufficiently justified the requested access, e.g., by showing necessity, and (2) that production-side protections, such as trade secrecy, shield the producing party from requests, regardless of the requesting party's justification.
Together, these two categories of arguments lead to a ``Catch-22'': that the requesting party cannot substantiate a request for access without supporting information, and the producing party is not compelled to grant access without further substantiation.
We term these dynamics the {``privatization of proof'': 
when control over proof increasingly falls in the hands of private parties that can 
demand justification for access while ensuring that justification remains out of reach.
}

In \Cref{sec:asymmetries}, 
we identify seven sources of asymmetry between parties in AI-related cases that contribute to proof privatization.
We additionally make two observations that will motivate our framework in \Cref{sec:test}:
\begin{enumerate}[left=8pt,topsep=2pt,parsep=0pt,partopsep=0pt]
    \item While access to models, 
    documentation, data, and user logs have received the most attention in ``AI access'' conversations, {\bf access to expertise, compute, and infrastructure are critical} to AI-related disputes.
    \item We argue that access is multidimensional and different forms of access are often {exchangeable}. 
    {\bf That access types are fungible allows substitutions that preserve a requesting party's ability to gather evidence while allowing the producing party to protect artifacts that it views as sensitive or confidential}.
\end{enumerate}
In \Cref{sec:test}, 
we propose a three-part test to guide courts in evaluating disputes over access to AI-related evidence. 
The test is intended as a practical tool for judges, as courts routinely use tests to improve consistency and predictability in the handling of disputes. AI-related cases frequently reach an impasse over access to evidence, addressing them in an ad hoc manner.
Our test draws on existing doctrines of relevance, proportionality, and necessity to create a  framework for evaluating access requests in AI-related cases.
The test is intended for pretrial disputes where the requesting party must make a threshold showing necessary for the claim to proceed.

\section{Analysis of AI, algorithmic, and automated technology cases}
\label{sec:strategies}

In this section, we describe common  arguments from our research that the producing party raises in response to requesting party's efforts to access evidence uniquely within producing party's control in cases concerning automated and AI technologies. We observe two recurring categories of arguments: (1) arguments that the requesting party has failed to show that access to the requested evidence is necessary to prove their claim, 
and (2) arguments that invoke production-side limitations on disclosure of evidence, independent of the requesting party's showing of need.
We find that the interaction between these two categories results in an \textbf{``Catch-22'': that the requesting party cannot substantiate a request for access without access to further evidence, and the producing party is not compelled to grant access without further substantiation.}
\begin{table}[htbp]
\centering
\caption{Recurring Arguments Against Evidentiary Access}
\label{tab:access-arguments}

\footnotesize
\setlength{\tabcolsep}{4.2pt}
\renewcommand{\arraystretch}{1.03}

\begin{tabularx}{\textwidth}{@{}
>{\raggedright\arraybackslash}p{0.28\textwidth}
>{\raggedright\arraybackslash}X
>{\raggedright\arraybackslash}p{0.25\textwidth}
@{}}
\toprule
\textbf{Argument} & \textbf{Access can be contested by arguing that...} & \textbf{Illustrative cases} \\
\midrule

\multicolumn{3}{@{}l}{\textsc{Request Side Arguments}} \\
\midrule

\textbf{1. Methods sufficient} &
High-level methodological descriptions or documentation adequately characterize the system, making implementation-level access unnecessary to prove the claim. &
\emph{People v. Superior Ct. (Chubbs)}, 2015; \emph{State v. Loomis}, 2016 \\

\addlinespace[0.2em]

\textbf{2. Non-independent expert testimony or evaluation credited} &
Developer-generated expert testimony or validation studies satisfy evidentiary demands, foreclosing independent evaluation despite potential conflicts of interest. &
\emph{People v. Superior Ct. (Chubbs)}, 2015; \emph{State v. Loomis}, 2016; \emph{State v. Pickett}, 2021 \\

\addlinespace[0.2em]

\textbf{3. AI role or scope of use disputed} &
Human decision-makers retained final discretion, the AI system played only a secondary or peripheral role, or it was not used in the specific workflow that affected the plaintiff. &
\emph{Mehrara v. Canada}, 2024; \emph{Huskey v. State Farm}, 2023; \emph{Est. of Lokken v. UnitedHealth}, 2025 \\

\addlinespace[0.2em]

\textbf{9. Model multiplicity} &
Output- or outcome-level access is insufficient because many plausible explanations may produce the same observed outputs, but only some would bear on liability. &
\emph{State v. Loomis}, 2016; \emph{State v. Pickett}, 2021 \\

\addlinespace[0.2em]

\textbf{10. Independent performance evaluation} &
The requesting party lacks sufficient access to evaluate system performance, group-level effects, and risks across relevant populations and deployment contexts. &
\emph{People v. Superior Ct. (Chubbs)}, 2015; \emph{State v. Loomis}, 2016; \emph{State v. Pickett}, 2021 \\

\midrule
\multicolumn{3}{@{}l}{\textsc{Producing Side Arguments}} \\
\midrule

\textbf{4. Proprietary information} &
Models, data, or documentation constitute competitively sensitive proprietary assets, including trade secrets. &
\emph{NYT v. Microsoft/OpenAI}, 2023--; \emph{Kadrey v. Meta}, 2023--; \emph{State v. Loomis}, 2016 \\

\addlinespace[0.2em]

\textbf{5. User or third-party privacy} &
User or third-party privacy interests block access to user logs or data relevant to the claim. &
\emph{NYT v. Microsoft/OpenAI}, 2023-- \\

\addlinespace[0.2em]

\textbf{6. Technical infeasibility or burden} &
Requested access is prohibitively costly, technically complex, or impossible to provide. &
\emph{Kadrey v. Meta}, 2023--; \emph{NYT v. Microsoft/OpenAI}, 2023-- \\

\addlinespace[0.2em]

\textbf{7. Alternative access formats/remedies and cost bearing} &
Disputes concern the format of access, remedies for lack of access, and allocation of associated costs. &
\emph{NYT v. Microsoft/OpenAI}, 2023--; \emph{Kadrey v. Meta}, 2023-- \\ 

\addlinespace[0.2em]

\textbf{8. Retention and logging procedures} &
Obligations around whether training data, model versions, output logs, or intermediate artifacts were retained, deleted, or must be preserved going forward are contested. &
\emph{NYT v. Microsoft/OpenAI}, 2023-- \\

\bottomrule
\end{tabularx}

\vspace{0.25em}
\begin{minipage}{\textwidth}
\footnotesize
\emph{Note:} Numbers 7--10 are discussed in the case review in Section B of the Appendix.
\end{minipage}

\end{table}

{\subsection{Case Research Methodology}
We researched relevant case law involving litigation of AI and similar automated decision-making systems to identify the recurring arguments described in this section. In Appendix \ref{sec:case_research}, we provide a detailed description of our case research methodology, case analysis for each case, and a case table tracking summary characteristics of the case and elements that support our central claims about the privatization of proof in AI litigation (Section \ref{sec:asymmetries}) and our proposed test (Section \ref{sec:test}), which follow this section.}

\subsection{Request-side Arguments Against Evidentiary Access} \label{sec:request_side_arguments}
Below, we discuss three types of arguments against evidentiary access that challenge the request itself.

\subsubsection{Producing parties argue that high-level descriptions or system outputs are sufficient, despite strong indications otherwise.}
Producing parties frequently challenge the necessity of implementation-level access by asserting that methodological information about the automated systems is sufficient to satisfy the evidentiary requirements of the requesting party's legal claim.

Courts have accepted this argument as a basis for denying discovery of source code in cases involving probabilistic DNA testing software. In \textit{People v. Superior Court (Chubbs)},\footnote{People v. Superior Ct. (Chubbs) No. B258569, 2015 WL 139069 (Cal. Ct. App. Jan. 9, 2015)}, the California Court of Appeals denied the criminal defendant's motion to compel discovery of the source code underlying TrueAllele, a probabilistic DNA analysis system, for a Kelly/Frye hearing challenging the reliability of expert testimony based on new scientific techniques. The court reasoned that the requesting party's demonstrated understanding of `TrueAllele's methodology, inferences, and reliance on the likelihood ratio'' in her declaration and publicly available materials describing the TrueAllele methodology, such as ``patent documents'' and ``published articles'', ``undercut'' the asserted need for implementation-level access.\footnote{\textit{Id}. at *8}
This reasoning {\bf rests on the assumption that methodological descriptions supply the minimum evidence necessary for the requesting party to prove its claims about the contested system.} Methodologies typically specify a class of plausible implementations, not the particular system as deployed. Legally relevant differences often arise at the level of implementation-specific details that have direct bearing on the legal theory advanced by the requesting party. In probabilistic, AI systems, this includes, but is not limited to, data processing choices, hyperparameter selections, or the setting of thresholds values.
Stated differently, AI systems are probabilistic (meaning that different outcomes can result from exactly the same inputs and also underdetermined (meaning that different systems can arise from nearly identical procedures). 
The result is: (i) descriptions of systems are often not sufficient to reproducing a system's errors or harms, and (2) while the description of a system may seem benign, it is still possible for the realized system to be harmful. 

In contrast, the Superior Court of New Jersey granted implementation-level access to TrueAllele in \textit{State v. Pickett}.\footnote{State v. Pickett, 466 N.J. Super. 270 (App. Div. 2021)}
We note that, for the court to reach this conclusion, the requesting party's made arguments about FST, a probabilistic DNA system based on the same method at TrueAllele. An audit of the FST source code revealed an implementation-level design choice in the system to exclude weakly informative loci, a choice not apparent from examining the methodology or documentation, which was later shown to overestimate likelihood of guilt.\footnote{\textit{Id}. at 307-308} Without this external system and the ability to audit its source code, such a determination would not have been possible. It is notable that that the source code for FST was only available due to a court order compeling disclosure in prior litigation of that system.\footnote{Order granting request to produce the source code underlying FST, United States v. Johnson, No. 15-CR-565 (VEC) (S.D.N.Y. June 7, 2016), ECF No. 57.}

\subsubsection{Producing parties advance unverifiable claims about system behavior, leaving courts reliant on potentially biased or non-representative evaluations in place of independent evaluations.}
Producing parties object to requests for access enabling independent evaluation by {\bf contending that expert testimony or existing validation studies (e.g., academic papers) provide sufficient evidence} for the requesting party to prove their claims about AI systems. 
When a propriety AI system is controlled by a developer with financial or litigation incentives aligned with the producing party, developer-generated evidence about the system may reflect those interests. Access enabling independent evaluation may be necessary for the requesting party to rebut potentially biased evidence about the system.

In \textit{Chubbs}, the court denied the defendant's request for access enabling independent evaluation of TrueAllele, reasoning that existing expert testimony and peer-reviewed validation studies were sufficient for the requesting party to assess reliability. That evidence, however, was largely generated by Dr. Perlin, the system's developer, who had a financial interest in TrueAllele and testified as an expert for the prosecution. In \textit{Pickett}, the court challenged the assumption implicit in \textit{Chubbs} court's reasoning that access enabling independent evaluation was unnecessary when developer-generated evidence about the system was available, particularly when the developer was in exclusive control of the system and had incentives aligned with the adverse party. The court emphasized that six of the seven peer-reviewed validation studies relied upon were authored by Dr. Perlin himself, that the remaining study acknowledged his professional involvement, and that the publications expressly disclosed his financial conflict of interest. \textit{Pickett} recognized the circularity of relying on potentially biased developer-generated evidence to deny access to independent evaluation necessary to rebut such evidence. 

These positions and justifications could be informative of courts' responses to requests for deep access to AI systems, where the producing party controls the system and much of the evidence offered to explain or defend it, including public blogs and papers. 
Independent evaluation access may also enable the requesting party to test the system behavior on the specific populations, inputs, or decisions contexts relevant to the litigant's claim.  
A substantial body of work in machine learning has shown that model performance can vary significantly under distribution shift \cite{koh2021wilds, d2022underspecification}, underscoring why prior validation studies are often a poor substitute for access that enables case-specific evaluation.

\emph{Chubbs} and \emph{Pickett} both arose under Frye-based standards, which condition admissibility on general acceptance of the underlying science in the relevant community, yet reached opposite conclusions on source code access. We survey all 24 TrueAllele proceedings in which source code access was contested in Appendix~\ref{sec:trueallele_all}, finding that courts declined to compel disclosure in 21 of 24 cases and this pattern holds across varying evidentiary standards.

In \textit{State v. Loomis},\footnote{State v. Loomis, 371 Wis.2d 235 (Wis., 2016)} the Wisconsin Supreme Court held that the use  of the COMPAS risk assessment at sentencing did not violate the defendant's due process right to be sentenced based on accurate information. Although the defendant could not review and challenge how the proprietary algorithm calculated risk, the court concluded that due process was satisfied because he could review and challenge the resulting risk scores in the presentence investigation report (PSI), he could review the algorithm's inputs drawn from his criminal history and questionnaire responses, and prior validation studies had determined the system to be reasonably accurate. \textit{Loomis} provides a useful context for examining the limits of treating access to a litigant's own input-output observation and general validation studies  as sufficient system access for the litigant to meaningfully contest a system's use in his case. Individual input-output observations or validation studies conducted on different reference populations or that do not isolate the defendant's class cannot distinguish whether an outcome is driven by individualized factors specific to the defendant, group-level effects, or a mismatch between the system's reference population and the actual decision context. The \textit{Loomis} court explicitly recognizes these concerns by requiring any PSI containing a COMPAS score to include a written advisement detailing risks associated with the system use including its proprietary nature, that risk scores are based on aggregate group data, the lack of validation studies on the Wisconsin population, and the need for ongoing re-norming as populations change.\footnote{\textit{Id}. at 276} Yet, despite acknowledging these risks, the court's analysis ultimately stops short of requiring access that would enable litigants to independent evaluate whether those risks are implicated in their own cases and contest the systems use on that basis.

\subsubsection{Producing parties question the relevance of the system to the disputed fact, undermining the basis for the access request.}
One way defendants have sought to deny access to AI systems is by classifying the technology as implementation or organizational tools rather than decision makers. In \textit{Mehrara v. Canada},\footnote{Mehrara v. Canada, [2024] F.C. 1554 (Can.)} Canadian authorities defended the use of a technology, Chinook 3+, within their immigration process. IRCC frames Chinook 3+ as merely a user interface, providing a ``visual representation of a client’s information'' to immigration officers responsible for determining visa outcomes. While Chinook itself is not an algorithm, it does display applicants' information in a way that can systematically influence officer's discretion, including a risk score predicted by the algorithm ITAT \cite{canadaChinookCIMM2022}. Another efficiency feature allowed officers to select amongst pre-selected reasons for denying individual's visa applications. %

The framing had direct procedural consequences. The court did not grant access to Chinook spreadsheets, officers' working notes, or the ITAT risk indicator score. Although the spreadsheets were deleted daily, the latter two were not found relevant. No risk score was given to this applicant, so the plaintiffs could not investigate how ITAT is portrayed or whether the lack of a score affects outcomes. Moreover, because Chinook was {\bf treated as an organizational tool, its internal presentation of information (such as the officer notes) was treated as immaterial to the outcome}. As a result, the claimant could not investigate whether the truncation or structuring of evidence might have constrained the officer's reasoning in practice. By defining the tool as a non-decisional interface, the informational environment that structured the officers' decision making and refusal is deemed irrelevant, and procedural scrutiny narrowed to the final judgement. The officer's reasoning, the court ruled, was unfettered by the technology.

However, Mehrara's counsel questioned whether discretion meaningfully remains independent under such systems: ``Is the officer really making a decision based on an application themselves, or...is the discretion being fettered by having this truncated single row of evidence for them'' \cite{AItriaging2025}? In later litigation, she described how Chinook's notes-generation interface allows officers to select standardized reasons for refusal. Notably, if their decision to deny an applicant is based on sufficiency of evidence, the officer may simply refuse the application, but if their concern is around credibility (that an applicant is being truthful in the reason around their trip), they have to contact the applicant to investigate and express these concerns. Moayyed notes in another Chinook case, \textit{Jahanian v. Canada},\footnote{Jahanian v. Canada, [2024] F.C. 581 (Can.)} ``with the widescale adoption of Chinook and the uniform reasons generated, the language that typically signals that an officer was concerned about credibility is disappearing and is replaced by language that signals concerns about sufficiency of evidence"  \cite{ziaieMoayyed_Chinook2024}. Credibility concerns might be disguised as evidentiary ones, which bypass safeguards that might allow the applicant further avenues for appeal.

The same ``AI as secondary to human-decision makers" framing appears in private litigation. {Jacqueline Huskey and Riian Wynn, two Black homeowners insurance claimants, sued State Farm under the Fair Housing Act ``alleging that State Farm handled the claims of Black homeowner insurance policyholders with greater scrutiny because of their race" \cite{StateFarmAlgorithmBias}. Notably, the Complaint cites a 2021 survey which shows ``white homeowners were almost a third more likely than Black homeowners to have their claim processed expeditiously (in less than a month), while Black policyholders were 39\% more likely to have to submit extra paperwork to justify their claims, causing months of delay in receiving coverage for urgent repairs" \cite{StateFarmAlgorithmBias}.

For a disparate impact claim under the FHA, Plaintiffs do not need to prove intent, merely that a policy produces unequal outcomes. However, to continue as a class action, they need to show that ``there are questions of law or fact common to the class" \cite{frcp23}. In short, there needs to be a unifying policy or practice that would allow all of the class action claims, which generally garner more strength through their numbers, to be resolved together.

State Farm admitted to using ``almost two dozen computer-based, automated and/or rules-based tools" for routing claims and assisting in identifying ``potentially fraudulent claims," but argued that claims were ultimately handled at the discretion of employees on an individual basis \cite{HuskeyStateFarmDocket}. The "role of the tool" is important, because State Farm calls upon the standard for class certification established in \emph{Wal-Mart Stores, Inc. v. Dukes} (2011).\footnote{\emph{Wal-Mart Stores, Inc. v. Dukes}, 564 U.S. 338 (2011).} This ruling establishes that individual discretion is not enough to bind a class together. In short, if State Farm is able to successfully argue that claims were handled at at the discretion of individual human adjudicators, rather than one unified algorithm, they can force Huskey and Wynn to pursue their claims on an individual, rather than class action, basis.}

In \textit{The Estate of Gene B. Lokken v. UnitedHealth Group},\footnote{The Estate of Gene B. Lokken v. Unitedhealth Grp., Inc., No. 23-CV-3514 (JRT/SGE), 2025 WL 2607196 (D. Minn. Sept. 8, 2025).} UnitedHealth defends its nH predict model, which is used ``to determine whether Medicare Advantage patients should receive post-acute care" \cite{oscislawskiUnitedHealthAI2025}. Plaintiffs argue that the tool overruled physicians and often resulted in the denial of life-saving treatment, which lends itself to breaking company policy: physicians are the ones making decisions regarding patients' care. While United insisted that the algorithm only acted as a guide for physicians, the high error rate and the affected populace of patients treated with the involvement of nH predict made the ``AI as a guide" argument less believable. Public reporting indicates a roughly 90 percent error rate, with the affected populace being the elderly, who are more likely to either die before contesting the decisions or lack the technical or institutional knowledge to go about a contestation. United's insistence that AI acted as a guide sought to protect the model from scrutiny by placing responsibility on the acting physicians, but it was this very allocation of responsibility that would allow the breach of contract claims to proceed.

Unlike in \textit{Mehrara}, plaintiffs were able to proceed by grounding their claims in a breach of company policy and bad faith. United's policies promise patients that their healthcare decisions be made by a physician. The plaintiffs argued that the deal was made in bad faith. Even if an acting physician did have nominal authority, the substance of the commitments were unmet given the model's deterministic nature.

\subsection{Production-side Arguments Against Evidentiary Access} \label{sec:production_side_arguments}
In contrast to production-side arguments that contest the necessity of the requested evidence, producing parties also resist evidentiary access by appealing to constraints internal to production. This division tracks the two elements of the standard for discovery in federal courts, relevancy and proportionality.\footnote{Fed. R. Civ. P. 26(b)(1)} Where the arguments in Section \ref{sec:request_side_arguments} seek to defeat discovery by challenging the value of the evidence to the requesting party, the latter emphasizes the cost of producing the evidence to the producing party.

\subsubsection{Private actors argue that access poses risk to competitive advantage, appealing to business confidentiality and trade secrecy.}
The \textit{New York Times Co. v. Microsoft Corp.}\footnote{New York Times Co. v. Microsoft Corp., No. 23-cv-11195 (S.D.N.Y. filed Dec. 27, 2023)} case illustrates how producing parties frequently resist evidentiary access by framing requested materials as competitively sensitive assets. At the onset of the case, the lead legal counsel denied access to the model's proprietary model information ``on commercial grounds,'' an argument that echoes across other AI copyright cases \cite{nytOpenAILawsuitComplaint2023}. In \textit{Kadrey v. Meta Platforms, Inc.}\footnote{Kadrey v. Meta Platforms, Inc., No. 23-cv-03417 (N.D. Cal. filed July 7, 2023)}, Meta explained that Llama 2's ``data mixes are intentionally withheld for competitive reasons” \cite{kadreyMetaComplaint2024}. Meanwhile, plaintiffs alleged that this framing works to ``avoid scrutiny by those whose copyrighted works were copied and ingested during the training process for Llama 2.'' Plaintiffs are forced to resort to proxy forms of proof: inference from public datasets, or, as the NYT did in it's Exhibit F, random querying of parts of its copyright work to induce a record of the model regurgitating its copyrighted works \cite{ExhibitF1_2023}. 
\textbf{Competitive advantage arguments can work to gatekeep and slow access to training data or source code}, rather than allow a mediated understanding of what level of access is appropriately proportional.

\subsubsection{Third-party privacy and confidentiality}
\textbf{Model providers have also sought to block access by appealing to users' privacy concerns.} In \textit{New York Times Co.}, Sam Altman met the request from plaintiffs for access to users' output data by invoking privacy concerns \cite{perez2025_altman_confidentiality}. Output data is critical to plaintiffs for a number of reasons. First, output data can provide direct evidence that the model itself reproduces copyrighted works. Second, even when courts conceptualize the user as the infringer, OpenAI can be linked to contributing to infringement. Patterns of contributory behaviour that might be supported by output data include the frequency of text regurgitation across the range of minimally to maximally infringing users queries. As of June 2025, the court has ordered OpenAI to retain their output logs, even when users have deleted the chats on their end. OpenAI is appealing the decision, on grounds that the decision infringes upon users' privacy. Sam Altman, in response, publicly called for something akin to ``AI privilege" to protect users \cite{perez2025_altman_confidentiality}. While privacy concerns have recently been invoked to block evidence, OpenAI itself has historically used user queries and output data as training data \cite{king2025userprivacylargelanguage}.

\subsubsection{Access is framed as infeasible or unduly burdensome to produce.}
{AI model providers and developers might also claim that providing access to, e.g., training data, is too technically difficult or cost intensive.} 
In \textit{Kadrey v. Meta Platforms, Inc.}, for example, the court ruled limited discovery to post-training data, as the raw dataset was ``massive compared to the datasets actually used.'' Legal analysts note the training data size can be used ``to [defendants'] advantage. Specifically, courts may be sympathetic to proportionality arguments when the data is burdensome to produce'' \cite{sternekessler2025trainingdata}.
If courts insist upon training data identification, there is often debate over who has to go through the identification process of relevant scraped works in the training data and by what degree of fineness. For example, in \textit{New York Times Co. v. Microsoft Corp.}, the matter of who is primarily responsible for this article identification remained a central point of contention. Initially, OpenAI built the infrastructure for the news plaintiffs to do this investigation themselves. OpenAI provided ``two virtual machines with computing resources. These machines were provided so that the counsel for NYT and Daily News could perform searches for their copyrighted content in its training sets” \cite{NYTevidence}. While the virtual machines are framed as granting access to plaintiffs, the sandbox structure externalized the technical labor, error risk, and verification burden onto plaintiffs despite OpenAI's exclusive control and higher familiarity with the training data. 

These virtual machines, or the ``sandbox,'' as it has been referred to in the case, came with several technical issues. First, mid-way through the process, “all of News Plaintiffs' programs and search result data stored on one of the dedicated virtual machines” got deleted \cite{helmer_eDiscoveryDay2024}. NYT alleged this deletion occurred at the hands of OpenAI engineers, while OpenAI claims the deletion was the result of improper handling of the sandbox. Regardless of where the fault lies, once the query information was deleted, the retrieved information was ``unreliable and [could not] be used to determine where the News Plaintiffs' copied articles were used to build Defendants' models” \cite{arsTechnica_OpenAIcourtTech2024}. 

The NYT explained they had spent over 150 hours gathering the evidence, and that the deletion had made it clear that OpenAI was best positioned to identify the data \cite{NYT_v_Microsoft_ECF328}. After the deletion, the news' plaintiffs filed an expansive Request for Admissions. 
In response, OpenAI claimed that the request was so burdensome that it would be unreasonable for them to comply: “Plaintiffs need to engage with OpenAI and meet and confer before seeking a court order compelling an answer to nearly 500 million requests for admission” \cite{500_mil_NYT_v_Microsoft_ECF305}. At present moment, OpenAI has agreed to bear further responsibility in helping the news' plaintiffs conduct the training corpus search, but the news' plaintiffs must provide more curtailed requests. 

It seems that the Court has come to a resolution with regards to how the training data will be identified using two methods: (1) a six-word n-gram token matching and (2) URL based matching. This means that the News plaintiffs have specific baseline URLS (such as www.nytimes.com) for their articles and sets of six-word sequences drawn from each work, which OpenAI can then search against its training data. However, this method might undercapture infringement in certain scenarios, such as when works are republished on secondary websites, stripped of their original URLS, paraphrased, partially quoted, or excerpted in less the chosen n-gram length. While the argument that evidence identification is too technically infeasible might have limited plaintiffs' requests for admission and prevented a higher level of access (say, for example, any time a set of three words is regurgitated), it may be litigiously advantageous for plaintiffs to shift the burden of identification to model providers.

Together, these developments indicate that AI developers and deployers can \textbf{argue that requests for access are infeasible or unduly burdensome, often taking advantage of the court's lack of technical knowledge}. Yet requesting parties can potentially push courts to make AI developers and deployers carry a greater burden, especially with respect to their own systems.

\section{Proof privatization, Systematic Asymmetries, and Fungible Access}
\label{sec:asymmetries}

In the previous section, 
we analyze arguments that producting parties may use to resist the requesting party's efforts to obtain evidence that is within producing party's control.
In this section, 
we study patterns across these cases. 

We first discuss the phenomenon of {\bf ``proof privatization''}: when control over proof falls in the hands of a party that can demand justification for access while ensuring that justification remains out of reach.
While key proof falling in the hands of reluctant private actors is not new, 
we discuss how seven sources of asymmetry between parties in AI-related cases can contribute to proof privatization.
We additionally make two arguments that motivate a test in \Cref{sec:test}. 
First, while significant attention has been paid to access to models, 
documentation, and recently data and user logs,
{\bf access to expertise, compute, and infrastructure are also critical} to disputing AI-related claims.
Second, access is {\bf multidimensional}; specifically, different forms of access are often \emph{\bf functionally exchangeable}. That access types are exchangeable allows for {\bf access substitutions} that preserve a requesting party's ability to gather evidence while allowing the producing party to protect sensitive or confidential artifacts. %
This observation is key to our paper as it implies that disputes over access do not need to collapse to a single option; rather, both parties have a range of options and can therefore engage in a nuanced, fact-specific negotiation of access.

\subsection{Proof Privatization}\label{ssec:proof-privatization}

That key information lies in the hands of one party of a dispute and not the other is not unique to AI. 
Litigation has long dealt with related issues, including trade secrecy complicating discovery; relevant evidence being lost or inadequately preserved; courts limiting discovery based on the cost of production; ``repeat players'' having procedural advantages; and technically sophisticated defendants understanding the subject matter far better than those challenging them.
Producing parties are often intentionally difficult, dragging their feet through strategies like ``document dumping''. The question, then, is whether AI-related cases present barriers to evidence that require special consideration.

In \Cref{sec:strategies}, we observe a Catch-22 that can determine outcomes:  the requesting party cannot substantiate a request for access without supporting information, and the producing party is not compelled to grant access without further substantiation.
As a result, litigation can fail due to what we call ``proof privatization'': 
when control over proof falls in the hands of a private party that can
demand justification for access while ensuring that justification remains out of reach,
exacerbated by the concentration of capital, resources, infrastructure, and expertise within the hands of private actors.
While this concentration of control has implications across multiple stages of litigation, our case analysis finds that it arises during the pre-trial stage, and the court's response often determines whether cases stall or settle.

Proof privatization risks undermining private litigation as a functioning mechanism for AI accountability. The problem extends beyond obtaining information: it affects the ability to support or refute claims, and to reproduce or debunk evidence. 
 In practice, disputes often reach an impasse in which the requesting party asserts that access is relevant and necessary, the producing party insists it is neither and that the materials are protected. 

\subsection{Seven Sources of Asymmetry}\label{ssec:seven-asymmetries}

Across contexts, there are seven sources of asymmetry between private actors (namely, AI developers and deployers) and the requesting party. 
\begin{enumerate}
\item \textbf{Model.}
This includes access to all or a subset of the model weights and/or activations, access to the model's outputs, and API access \citep{casper2024black,cen2024transparency}. Model access is often critical because it allows direct testing of the system of interest on different inputs. Models are highly expensive to train, so replicating models is costly.

\item \textbf{Data.}
This includes pre-training, supervised fine-tuning, human feedback, and benchmark data. 
There are various forms of data access, including full database access, exact n-gram search, and fuzzy search \cite{shenfallacies,panda2025privacy}.
Data is highly expensive to gather, considered proprietary, or protected as private \cite{lee2023explainers,lee2023talkin}.

\item \textbf{Documentation.}
 We refer to documentation as information that reveals design decisions and organizational knowledge. In practice, documentation provides critical insights that are difficult to infer, but it is therefore the most revealing and thus generally considered to be highly confidential \cite{sedona2022tradesecrets}.

\item \textbf{Post-deployment information (logs and operational traces).}
Post-deployment traces serve as the primary source of information of how a system behaves and is used in practice. Recent litigation illustrates that the availability and retention of such logs can become a contested issue \cite{reuters2025openai_preserve,nytorder2025preserve}.

\item \textbf{Expertise.}
AI development and deployment as well as testing and evaluation generally require specialized knowledge and experise.  Expertise also shapes whether a party can formulate targeted discovery requests and assess (or rebut) technical claims that factfinders may otherwise credit \citep{drcf2022auditing}.\footnote{We note that expertise can factor in multiple ways. Here, we refer to AI expertise. Other forms of expertise, such as business or legal expertise, can speak to relevant decisions. We do not focus on these forms of expertise, as they are not specific to AI-related cases.}

\item \textbf{Compute.}
 Many operations---including red teaming and reproducing a model via training---require significant compute. Limited compute can therefore impose limits on what analyses are feasible \citep{ainow2018litigating_algorithms,cen2025audits_resource_constraints}.

\item \textbf{Infrastructure.}
 Similarly to compute, AI developers and developers often have access to tooling that greatly assists in AI development and analyses. While lack of access to infrastructure is not necessarily prohibitive, it can create significant barriers for external parties \citep{ojewale2025audittooling}.
\end{enumerate}
We conclude with three takeaways.
First, while debates over asymmetry in AI disputes have often centered on \emph{model access} and increasingly on \emph{data access}, they are not determinative of how a system behaves. 
Second, {one party in a suit typically has significantly greater access than the other to all artifacts above.} 
Third, the last three asymmetries are discussed less often even though expertise, compute, and infrastructure are of utmost significant.

\subsection{Different Types of Access are Functionally Exchangeable}\label{subsec:exchangeability}

Access in AI disputes is often treated as a one-dimensional question: greater or less access to a particular artifact such as {the} model, training data, or  source code. One of our central observations is that access is \emph{multidimensional}; specifically, most types of access  can be \emph{functionally exchangeable} with other types of access. Because they often serve similar functions, denying one form of access does not always foreclose the successful gathering of evidence if other forms of access can be provided. For example, limited model access may be partly mitigated by richer documentation plus sufficient compute and infrastructure to run systematic tests; conversely, where documentation is contested, significant access to model, data, and logs may support many of the same inferences.

We build our main contribution in \Cref{sec:test} on this observation: exchangeability makes it possible to treat access not as an all-or-nothing entitlement to specific artifacts, but as a set of negotiable options oriented toward an evidentiary objective. This exchangeability matters most in disputes where the producing party claims heightened risk from disclosure (e.g., trade secrecy, privacy, security, or misuse) while the requesting party seeks access for a specific evidentiary purpose. Exchangeability allows one to ask: \emph{what access is sufficient to accomplish the requesting party's evidentiary task?} In many cases, that task can be satisfied through alternative combinations of access that pose materially different risks. For example, rather than producing sensitive datasets in full, a developer might provide structured evaluation sets, statistically representative slices, or controlled log access that enables the requesting party to test the contested proposition. Rather than disclosing full source code, the producing party might provide design and evaluation documentation, model cards, internal test results, or supervised testing access calibrated to the claim at issue.

This flexibility also helps explain why substitutions may be attractive even when they are not ``cheaper'' in any absolute sense. If access $A$ is especially valuable to the requesting party but uniquely sensitive to the producing party, the producing party may rationally prefer to offer access $B$ that is more costly (or less convenient) for the requester to use but materially less risky to disclose.
Thus, substitutions can protect what the producing party is most concerned about while still providing key access to the requesting party.

\section{Proposal: Three-part test}
\label{sec:test}
In this section, we build on our analysis and propose a three-part test that can be used to guide courts in resolving disputes over access to AI-related evidence. 
As discussed above, AI-related cases frequently reach an impasse over access, and this test removes reliance on ad hoc judgments of the arguments.
Our test does not create new procedural rules; rather it draws on existing doctrines of relevance, proportionality, necessity, and cost-shifting to give judges a structured framework for assessing whether proof privatization has occurred, to what extent, and whether requested access is reasonable.
The test should apply primarily in pretrial disputes, when a court must decide whether to allow discovery, inspection, testing, expert access, or similar evidentiary measures needed for a party to make a threshold showing. In practice, this includes Rule 26 discovery disputes, motions to compel or for protective orders, and pretrial admissibility or expert-access challenges. It may also guide later rulings if access to AI-related evidence remains necessary to fairly resolve the case and that evidence is substantially controlled by the opposing party.
The Zubulake factors discussed in \Cref{ssec:justification} for electronic discovery offer a useful analogy; in that case, a court organized existing principles into a reusable framework without formal rulemaking.
Finally, we note that a key feature of our test is that it accommodates varying standards by assessing proportionality relative to a flexible baseline.

\subsection{Three-Part Test}\label{ssec:test}

We refer to the requesting party as the party seeking access in order to support a claim, 
and the producing party as the party that controls an artifact to which the requesting party is requesting access.
The producing party may refer to multiple parties that, together, hold control over artifacts that the requesting party seeks.

\textbf{Step 1: Degree of asymmetry.} The first part of the test evaluates the degree of asymmetry between the requesting party and producing party, determining whether obtaining more access is necessary and relevant. 

    Mirroring the asymmetries given in Section \ref{sec:asymmetries}, this step can be performed by evaluating each party's access to relevant (1) models, (2) data, (3) documentation, (4) compute, (5) infrastructure, (6) expertise, and (7) downstream information. 
    Along all seven dimensions, the test should separately assess, at the time of evaluation,  what artifacts the requesting party and producing party have access to. 
    At this stage, the test should terminate if either one of the following holds:
    (i) the requesting party should be able to prove their claim with their current level of access, 
    or (ii) the requesting party would not be able to prove their claim even with the same amount of access as the producing party.
    The checks (i) and (ii) provide  indications of the {\bf necessity} and {\bf relevance}, respectively, of granting more access. 
    The test only continues if the requesting party {\em cannot} prove their claim with the current level of access, and the requesting party {\em could} prove their claim if they were in the producing party's position.

    \textbf{Step 2: Justified access.} 
    The second part of the test determines whether privatization of proof has occurred, and, in doing so, creates a standard for proportionality in cases of proof privatization in AI-related cases. 
    Proof privatization occurs when access to key artifacts is protected by the producing party.
    A critical choice in assessing proof privatization is establishing an appropriate notion of ``key'' evidence. 
    We propose that it should be determined by the \emph{minimal access needed to reproduce the elements of the cause of action}.
    For example, in a discrimination case based on observations of a systematic imbalance in hiring and knowledge that an automated ranking system is used, ``minimal access needed to reproduce elements of the cause of action'' corresponds to what minimal information and access to the automated system would, together, reproduce the systematic imbalance. In this case, query access to the system, information on how the employer uses the system (e.g., the input prompt), and possibly a deanonymized sample of applications may be needed.
    Specifically, the test should assess the quantities:
    \begin{enumerate}
        \item \textsc{Requested Access Benefit}: across the seven types of access, the benefit to the requesting party that the requested level of access would provide. 
        \item \textsc{Requested Access Risk}: across the seven types of access, the risk that the producing party would face if they were to grant the requested level of access.
        \item \textsc{Cause of Action Access Benefit}: how much benefit access equivalent to the minimal access needed to reproduce the elements of the cause of action would provide to the requester. 
        \item \textsc{Cause of Action Access Risk}: the risk that the producing party faces by granting the requester the minimal access needed to reproduce the elements of the cause of action.
    \end{enumerate}
    These could be measured monetarily or otherwise, as long as the units are consistent.
    Then, the benefit-to-risk ratio of the requested access (or cause of action) is the benefit of the requested access to the requesting party relative to the risk the producing party faces by granting the requested access (or minimal access to reproduce elements of the cause of action).
    If 
    \begin{align*}
        \frac{\textsc{Requested Access Benefit}}{\textsc{Requested Access Risk}}
        \geq
        \frac{\textsc{Cause of Action Access Benefit}}{\textsc{Cause of Action Access Risk}} ,
    \end{align*} 
    the benefit-to-risk ratio of the requested access is greater than the benefit-to-risk ratio of the cause of action access, 
    and the test favors the requested access.
    Otherwise, the test terminates.
    By comparing the requested access to the minimum access needed to reproduce the elements of the cause of action, the test ensures that the requesting party is entitled to some access and {\em anchors it at the cause of action}. Because the law already recognizes the cause of action as a legitimate basis for a claim, it provides a flexible yet acceptable threshold that is applicable across types of claims.
    It is also minimal in scope, as only gives the requesting party access to artifacts that are directly pertinent to the original cause of action.

 \textbf{Step 3: Alternatives and safeguards.} 
   The test only reaches this point if the request for access has passed Step 2.
     However, the producing party has one last defense that they can use to reduce or modify the access that requesting parties are granted. 
     The producing party can do so by proposing alternative plans for access,
     including a different combination of artifacts, conditions under which access is granted, the use of a sandbox, phased discovery, and other protective measures. 
      They could even propose to permit adverse inference instead of granting access if the producing party truly believes that  confidentiality is paramount and no amount of access protects these interests.
     An alternative is acceptable only if (i) it provides functionally equivalent access with respect to what the requesting party seeks to prove and (ii) it does not decrease the benefit-to-risk ratio of the requested access as assessed in Step 2.
     In other words, it cannot result in functionally less access that what Step 2 grants the requesting party, \emph{and} it cannot reduce the benefit-to-risk ratio accepted during Step 2.

\subsection{Justification}\label{ssec:justification}

\paragraph{A shared framework that applies existing doctrines to the problem of AI proof privatization.}
The three-part test that we propose respects existing procedural commitments and producing parties' confidentiality interests. US courts have long dealt with situations in which key evidence is held primarily or exclusively by one party. Doctrines governing discovery (the pretrial procedure through which parties obtain evidence), relevance (whether evidence bears on the disputed claim), proportionality (whether the burden of producing it is suitable relative to the needs of the case), cost-shifting (the allocation of discovery expenses between parties), spoliation (sanctions for the destruction of evidence), adverse inference (allowing the factfinder to draw unfavorable conclusions in the absence of evidence), and trade secret protection (safeguards limiting disclosure of sensitive business information) all reflect judicial efforts to manage informational asymmetries and the burdens of producing evidence \footnote{Fed. R. Civ. P. 26(b)(1), 26(c), 37(e); Oppenheimer Fund, Inc. v. Sanders, 437 U.S. 340, 351 (1978); Zubulake v. UBS Warburg LLC, 217 F.R.D. 309, 322--24 (S.D.N.Y. 2003); Residential Funding Corp. v. DeGeorge Fin. Corp., 306 F.3d 99, 108--09 (2d Cir. 2002); Seattle Times Co. v. Rhinehart, 467 U.S. 20, 34--36 (1984)}.

For example, \emph{Zubulake v.\ UBS Warburg LLC} was an employment-discrimination suit, where there was a request for emails for which many potentially responsive messages were stored on the defendant's backup tapes and would be costly to restore and search. The court used a seven-factor framework to assess whether to restore the emails, employing phased discovery and sampling to estimate the likely value of the information relative to the burden of production prior to ordering large-scale restoration. Ultimately, the court permitted restoration but employed cost-shifting, requiring the requesting party to bear a portion of the costs.
This and related cases demonstrate that courts recognize the risk that exclusive control over evidence can undermine the effective enforcement of substantive law, using tools like relevance, proportionality, cost-shifting, necessity, and cost-benefit analysis to assess the need to rebalance evidentiary burdens.

Yet existing doctrine addresses these issues in a fragmented manner. For example, discovery doctrine typically evaluates access through relevance and proportionality once litigation is underway, often assuming that requesting parties can reach the discovery stage without interrogating whether key evidence has been rendered inaccessible in the first place. Moreover, trade secret doctrine treats confidentiality as a competing interest to be balanced against disclosure, rather than as a structural feature of proof. 
As a result, courts may reach inconsistent outcomes.
{\bf The proposed test draws on these familiar doctrines, consolidating them into a shared framework specialized for AI-related disputes.}
We discuss specific design choices of the test below. 

\paragraph{Unpacking each step.}
The test provides a standardized way to assess the degree of proof privatization and the proportionality of the requested access while ensuring that the access granted is relevant, necessary, and minimal.

Step 1 formalizes whether parties face a meaningful informational asymmetry and whether additional access is both necessary and appropriate. 
It examines whether the requesting party cannot prove the claim with their current level of access \emph{and} whether the request for further access actually advances their ability to do so. 
If these initial conditions are not met, then the request for access is unwarranted and the test terminates. 

While Step 1 evaluates relevance of the request for access, 
Step 2 returns to the central problem: the degree of proof privatization.
Building on \Cref{sec:asymmetries}, it assesses the degree of asymmetry relative to the factors in \Cref{ssec:seven-asymmetries}.
The ``access benefit'' captures the amount the requesting party would expend to emulate the target amount of access. 
The access benefit can be very large in some cases, e.g., if the requesting party lacks access to large models and datasets that the producing party own, which may cost significant amounts to replicate. 
The ``access risk'' captures the risk that the producing party would face if they were to grant the requested access; notably, risk does not simply capture the possible amount the producing party could lose, but also the likelihood of such an outcome. 
For instance, if the producing party believes that there is a possibility that access could leak a trade secret, then risk captures not only the value of this trade secret but also the likelihood of such a leak.\footnote{The reason that the ``access benefit'' is not exactly analogous to risk in considering the chance of proving the claim with the target access is (i) the chance of proving a claim with the target access cannot be predetermined, as it central to the outcome of the suit, and (ii) Step 3 ensures that access is minimal and thus indirectly accounts for the possibility of over-estimating the numerator of the benefit-to-risk ratio.}

There are two components of Step 2 that are critical: (i) the use of the {\bf cause of action} as a baseline, and (ii) the {\bf comparison of ratios that effectively measures the proportionality} of the requested access.
Step 2 compares the access request to a legally grounded baseline: the \emph{minimal access needed to reproduce the elements of the cause of action}. 
Thus, it does not permit the requesting party maximal transparency, but ensures that some access is granted, lower bounded by what the requesting party should minimally be entitled in order to make (or contest) a claim the law already recognizes. 
Using cause of action anchors Step 2 at stable reference point to distinguish between excessive fishing expeditions and access that is minimal for adjudication.
Step 2 then compares the benefit-to-risk ratio of the requested access to the cause-of-action access. 
If the former exceeds the latter, then the requested access has a higher benefit-to-risk ratio than the baseline cause-of-action ratio, meaning that the test favors the requested access.
This test almost literally maps to a quantitative test of proportionality.
As sanity checks, note that the requester is not incentivized to request large amounts of access simply because the numerator of the benefit-to-risk ratio grows with greater access because the risk (the denominator) may also grow with greater access.
Finally, note that, as a bare minimum, the requester could request the minimal access needed to reproduce the elements of the cause of action, which would be accepted by Step 2.

Step 3 allows alternative proposals to ensure that the access granted is {\bf necessary and minimal} while  {\bf reducing risks to the producing party as much as possible given the amount of access granted}.
While Step 2 requires some consideration of the relative benefits and risks of requesting access, Step 3 is a critical safeguard that ensures that the producing party can propose alternatives or revisions to the request, as long as it does not contradict reverse the guarantees provided by Step 2.
Step 3 relies on the key observation from \Cref{sec:asymmetries} that \emph{access is not linear}: different forms of access can substitute for one another, meaning that if a producing party prefers to allow functionally equivalent (or more) access to the requesting party because they believe it poses less risk based on an internal calculation of their interests, then they should be permitted to do so.
The producing party may even wish to permit adverse inference instead of granting access if some form of confidentiality is absolutely paramount to them that they are willing to risk an adverse outcome with respect to the claim of interest in order to protect these interests.

\paragraph{Why the test does not add ``business confidentiality'' as an override in Step 3.}
The typical discussion of AI ``access'' revolves around the push-and-pull between the need for access and the desire to protect business confidentiality.
Such conversations often come to a standstill because claims of confidentiality are permitted to override the need for access.
Against this backdrop, we arrive at the issue of privatization of proof. 
Based on our analysis, our test focuses not on access versus business confidentiality, but on the relevance, proportionality, and necessity of the requested access.
In doing so, this test still \textbf{places significant weight on confidentiality interests while avoiding the pitfalls of a vague exception that can swallow access altogether}.

The framework assumes from the outset that the entitlement is bounded by what is minimally required to reproduce the elements of the claim. Confidentiality remains protected not via a vague exception, but via the combination of (i) an assessment of necessity and relevance in Step 1, (ii) a calculaton of risk to the producing party in Step 2, and (iii) the allowance of procedural safeguards, substitutions, and alternatives in Step 3. 

\subsection{Example}\label{sec:examples}

Suppose a publisher sues an LLM developer after preserving several outputs in which a model reproduced substantial portions of its paywalled articles. The developer responds that the outputs do not reflect copying of the publisher's works, but only ordinary prediction from similar reporting and predictable journalistic language. At Step 1, the asymmetry is substantial because the publisher has the articles and the regurgitated outputs, but the developer controls the relevant data, model, documentation, infrastructure, and user logs. 
The claim therefore survives Step 1 because the publisher already has a cognizable claim but cannot fully prove their copyright claim or rebut the developer's alternative explanation without additional access.

At Step 2, the court must compare the publisher's requested access against the cause-of-action baseline---that is, the minimum additional access needed to reproduce the copyright claim in a meaningful way.
In this scenario, reproducing elements of the cause of action implies using the LLM to systematically reproduce the publisher's articles, including articles that may not have been in the original claim, e.g., via query access to all relevant models. 
Importantly, even though the publisher's goal is to refute the claim that the LLM's output is due to ordinary prediction, the cause of action baseline is different.
However, the publisher may wish to ask for different access; under Step 2, they are permitted to do so as long as the access request has a larger benefit-to-risk ratio than the cause of action baseline. 
For example, the publisher may ask for search/query access to the pretraining data corpus; this may fail if the developer makes a strong case that the risk of data access poses a significant risk to their business, though the risk seems low (relative to the query access cause-of-action baseline above) since the pretraining corpus for modern-day large models is equivalent to a billion books of text.
On the other hand, access to the full pretraining corpus or model weights would likely fail Step 2 because their incremental benefit may be lower relative to their production-side risk.
A separate request for de-identified user-log data might also pass, not to prove copying into training, but to prove substitution or scale of dissemination. 

At Step 3, the developer may propose alternatives. 
For example, suppose the request is for query access to multiple models with the justification that if models of various size that were released \emph{before} an article was published cannot produce an article, but later models of comparable sizes that were released \emph{after} the article can produce it, then this is evidence that the regurgitation is not due to ordinary prediction.
However, suppose the developer argues that older models have been retired and running these deprecated models is not feasible or highly costly (a preservation issue). 
Then, the developer may prefer and thereby propose alternatives, such as limited search/query access to the training data corpus, 
such as URL or n-gram matching. 
By Step 3, those substitutes are acceptable only if they are functionally equivalent and do not reduce the benefit-to-risk ratio that Step 2 recognized. 
A second example is given in \Cref{sec:second_example}.

\section{Conclusion}

Although less often discussed than legislative or administrative actions, private ligitation is a key avenue for AI accountability.
For it to remain viable, parties bringing meritous claims about AI systems, decisions, and outcomes must be able to produce evidence that supports their claims. 
In this Article, we study the barriers AI ``challengers'' face when requesting access to key evidence. 
From our case analysis, we identify several key trends. 
For one, claims about automated, algorithmic, and AI technologies often fail during pre-trial stages, often due to a self-defeating cycle where the requesting party cannot gather clear evidence to support a request for discovery, but without discovery, they cannot produce evidence needed to move forward with the case. 
We pinpoint the underlying problem as one of ``proof privatization'': that major channels to and control over evidence are increasingly in the hands of private actors, shifting authority from the court to the private actors.
In response, we propose a three-part test for proof privatization in AI-related cases, building on existing legal principles.

\begin{acks}
We are indebted to Riana Pfefferkorn, Daniel E. Ho, Peter Henderson, Ziyaad Bhorat, and Nel Escher for their invaluable feedback and guidance. We also greatly appreciate several lawyers who wish to remain anonymous for their willingness to discuss their experiences with us. 
This work was supported in part by the Stanford Interdisciplinary Graduate Fellowship (SIGF) and the CSET Foundational Research Grants Program.
\end{acks}

\bibliographystyle{ACM-Reference-Format}
\bibliography{refs.bib}

\clearpage

\appendix

\section{Background and Related Work}\label{sec:evidence_bg}

\textbf{Information, resource, and expertise asymmetries.}
Many previous works emphasize that \emph{parties who control relevant information can shape what is knowable in litigation, particularly through secrecy claims and institutional deference to confidentiality}. Wexler describes how secrecy claims can block scrutiny in high-stakes contexts and how courts may accept those claims with limited balancing against constitutional rights \cite{wexler2018life}. 
In parallel, socio-legal scholarship examine how evidentiary and procedural rules reflect power and capacity, e.g., Galanter presents a repeat-player framework, which predicts that \emph{experienced, well-resourced parties are systematically advantaged} in navigating procedural demands and generating proof over time \citep{Galanter1974Haves}. Further works (often empirical) find that adverse outcomes often point to a limited capacity to assemble legally sufficient evidence---due to constraints of representation, time, and knowledge---rather than weak claims \citep{Sandefur2014LegalNeeds,AlbistonSandefur2013Access}. 
There is also documentation of how the high cost of legal services and unequal access to professional support shape who can meet proof requirements in practice \citep{Rhode2004Access,Hadfield2017Rules}. 
Finally, several works examine how \emph{expert testimony is strictly gatekept, with uneven consequences}. Work on expert evidence and post-{Daubert} gatekeeping emphasizes that reliability-focused admissibility doctrines influence who can realistically meet proof demands in complex litigation \citep{Faigman2001Daubert,Bernstein2014DaubertResistance}. Other works question why expert evidence is subject to heightened scrutiny,  underscoring how strict admissibility thresholds excludes relevant expert insights \citep{Schauer2010Expert}. 
\medskip 

\noindent 
\textbf{Procedural bottlenecks.}
Related works pinpoint unequal burdens at specific procedural moments. At the pleading stage, Dodson argues that plausibility pleading operates as an early evidentiary screen by implicitly demanding information often unavailable pre-discovery \citep{Dodson2010NewPleading}. 
Reinert argues that modern pleading standards create a procedural self-defeating cycle: plaintiffs are required to plead specific facts to unlock the doors to discovery, yet those very facts are often hidden within the defendant's private records; by demanding evidence this early, these standards systematically bar meritorious claims where the defendant holds all the information \citep{Reinert2011Costs}. 
Issacharoff and Miller critique the current motion-to-dismiss standard for allowing defendants to defeat a claim by highlighting a lack of evidence that only the defendants themselves possess. To fix this, they suggest rules that would require defendants to disclose key facts before they can seek a dismissal \citep{IssacharoffMiller2013Dismiss}. 

Although discovery doctrine is often framed as the response to informational asymmetry, it is also shaped by resource disparities.
Rowe argues that the rules governing who pays for the gathering of evidence are as important as the law itself, because if a party cannot afford the cost of discovery, they are effectively stripped of the ability to prove their case \citep{Rowe1982DiscoveryCost}.
Moreover, while discovery is a powerful mechanism for obtaining evidence, its scope is limited by Rule 26, which amended the definition of what is discoverable from anything that is ``reasonably calculated'' to lead to evidence to ``proportional to the needs of the case'' in order to prevent plaintiffs from going on costly ``fishing expeditions.''
Mullenix warns Rule 26 can protect poorer litigants from being outspent by their opponents, but  it can also provide wealthy defendants with a legal excuse to hide evidence by claiming it is too expensive to produce \cite{Mullenix2016Discovery}.

Another common critique is that discovery can be disproportional in providing too much evidence so as to raise legal costs \citep{lahav}. Thus, the proportionality test also exists to limit the amount of evidence provided. Still, there are instances in which too much evidence is provided to overwhelm legal opponents with information.
\medskip 

\noindent 
\textbf{Doctrinal responses: presumption, adverse inference, and spoliation.}
Finally, evidence law sometimes responds directly to systematic proof barriers with presumption and burden shifting. Doctrines such as \emph{res ipsa loquitur} relax production burdens where direct evidence is inaccessible and defendants are better positioned to explain underlying events \citep{Prosser1949ResIpsa}. Scholarship on presumptions emphasizes that burden allocation reflects judgments about access to information and fairness, not merely probabilistic inference \citep{Cleary1959Presumptions}. Related work on spoliation and Rule~37(e) examines how sanctions and adverse inferences attempt to rebalance proof burdens when evidence is lost or destroyed, while also highlighting the limits of remedial doctrine once information has disappeared \citep{Steinman2012Adverse,Koesel2017Spoliation}.
\medskip

\noindent 
\textbf{Discussions of evidence and transparency in AI governance.}
In parallel, debates over algorithmic governance and automated decision systems, scholars similarly argue that opacity and confidentiality can frustrate meaningful contestation and accountability \citep{pasquale2015black,CitronPasquale2014,citron2008_technological_due_process,desai2017trust,kroll2017_accountable_algorithms,blochwehba2020_access_to_algorithms}. Related work focuses specifically on code and forensic technologies, highlighting how intellectual property and trade secrecy claims can constrain testing and challenge, including in criminal justice settings \citep{katyal2019_paradox_source_code_secrecy,ryan2020_secret_algorithms,ryan2022_criminal_justice_secrets,siems2022_trade_secrecy_forensic_technology}.
Recent work emphasizes that resource inequality can also be infrastructural: compute, data access, and evaluation capacity are unevenly distributed, affecting who can generate and validate technical evidence \citep{ahmed2020de,casper2024black,cen2024transparency}.
When it comes to the scrutiny of expert testimony, concerns over gatekeeping expertise are increasingly salient in disputes involving technical systems, where parties may disagree not only about facts but about the methods required to establish them \citep{desai2017trust,kroll2017_accountable_algorithms}.
Aligned with Issacharoff and Miller's \citep{IssacharoffMiller2013Dismiss} critique, the current standards allow defendants to defeat claims by pointing to a lack of evidence that only the defendants themselves possess, scholars argue that these dynamics are amplified because key facts about design choices, training data, and validation practices are typically held by defendants \citep{kim2017datadriven}, producing a situation in which plaintiffs struggle to plead and prove discrimination without access to the very information that discovery is increasingly conditioned upon \citep{foggoVillasenor2021disparateImpactRule,bernt2021workplaceTransparency}.

\section{Case Research}
\label{sec:case_research}
\subsection{Case Research Methodology}
The authors surveyed a range of cases where plaintiffs sought to sue AI system developers or deployers under various legal theories, including: constitutional claims (due process, equal protection, and confrontation), federal statutory claims (disparate impact under Title VII and the Fair Housing Act, copyright infringement), and state common law claims (contract and tort). These cases span multiple procedural stages at which evidentiary access disputes arise, including pretrial admissibility hearings, motion to dismiss, discovery, summary judgment, and post-conviction challenges. Since most AI litigation has not been resolved to final judgment yet, we also researched earlier cases involving automated decisionmaking/predictive technology with analogical characteristics to present day AI. Because this area of law is relatively nascent and actively developing, there are few comprehensive traditional secondary sources that provide a jumping off point for relevant cases (e.g., legal treatises, casebooks). We compiled an initial list of cases drawing on significant cases based on the authors' existing knowledge, the \textit{Law and Algorithms} casebook \cite{sellars2024lawandalgorithms}, and news reporting on ongoing litigation. We expanded this initial list by examining cases cited by and citing to those cases. A further challenge was that many evidentiary access disputes do not appear in final published opinions but rather in interlocutory court orders that are often buried deep in litigation dockets and poorly indexed. Additionally, because many of the evidentiary admissibility challenges and contract/tort claims have proceeded in state courts, the available information is often fragmented across jurisdictions. To address these gaps, we supplemented our research with AI-assisted legal research tools, specifically Westlaw AI-Assisted Research and PrecedentPilot\footnote{\url{https://precedentpilot.com/}}, using semantic searches that began with broad queries (e.g., ``requested access to training data”) and narrowed progressively (e.g., ``citing user privacy concerns”) as our research and arguments developed.

\subsection{Case Overview}
We provide a high-level case overview table, Table \ref{tab:case_table}, summarizing key characteristics of each case and coded relevant elements implicated by each case, common arguments invoked to challenge evidentiary access (Section \ref{sec:strategies}) and features that give rise to access trade-offs motivating our three-part test (Section \ref{sec:test}). The coded elements are described in Table \ref{tab:code_table}. In Section \ref{sec:case_analysis}, we provide case analysis of each case researched, with more in-depth description of coded elements, and notable evidentiary access disputes and their resolutions by the court.

{\fontsize{7pt}{8pt}\selectfont
\begin{table}[h]
    \centering
    {\color{black}
    \begin{tabulary}{\textwidth}{LL}
         \toprule
         1 & \textbf{Methods sufficient} Producing party argues that high-level methodological descriptions or documentation adequately characterize the system, making implementation-level access unnecessary to prove the claim \\
         2 & \textbf{Non-independent expert testimony or evaluation credited} Producing party relies on developer-generated expert testimony or validation studies to satisfy evidentiary demands, foreclosing independent evaluation despite potential conflicts of interest \\
         3 & \textbf{AI role or scope of use disputed} Producing party argues that human decision-makers retained final discretion over outcomes, that the AI system played only a secondary or peripheral role, or that it was not used in the specific workflow that affected the plaintiff, narrowing the scope of discoverable materials \\
         4 & \textbf{Proprietary information} Producing party resists disclosure by asserting that models, data, or documentation constitutes competitively sensitive proprietary assets (e.g., protected as trade secrets) \\ 
         5 & \textbf{User or third-party privacy} Producing party invokes user or third-party privacy interests to block access to user logs or data relevant to the claim \\
         6 & \textbf{Technical infeasibility or burden} Producing party argues that requested access is prohibitively costly, technically complex, or impossible to provide \\
         7 & \textbf{Alternative access formats/remedies and cost bearing} Disputes over the format through which requested access is provided or remedies for lack of access and the cost burden on each party associated with the arrangement \\
         8 & \textbf{Retention and logging procedures} Contested obligations around whether training data, model versions, output logs, or intermediate artifacts were retained, deleted, or must be preserved on a forward-going basis \\
         9 & \textbf{Model multiplicity} When output or outcome-level access is insufficient to attribute or contest specific claims because many plausible explanations may produce the same observed outputs, but only some of these explanations would bear on liability \\
         10 & \textbf{Independent performance evaluation} Whether the requesting party has sufficient access to evaluate system performance, group-level effects, and risks across relevant populations and deployment contexts \\
         \bottomrule
    \end{tabulary}
    }
    \caption{Code Table}
    \label{tab:code_table}
\end{table}
}

\clearpage 

{\color{black}
{\fontsize{7pt}{8pt}\selectfont
\begin{longtable}{p{1.75cm}p{2cm}p{2.85cm}p{2.9cm}p{0.05cm}p{0.05cm}p{0.05cm}p{0.05cm}p{0.05cm}p{0.05cm}p{0.05cm}p{0.05cm}p{0.05cm}p{0.05cm}}
  \caption{Case Table} \label{tab:case_table} \\

  \toprule
  Case name & Legal theory & Evidence requested & Evidence obtained
  & 1 & 2 & 3 & 4 & 5 & 6 & 7 & 8 & 9 & 10 \\
  \midrule
  \endfirsthead

  \multicolumn{14}{l}{\tablename\ \thetable{} -- \textit{continued}} \\[0.5ex]
  \toprule
  Case name & Legal theory & Evidence requested & Evidence obtained
  & 1 & 2 & 3 & 4 & 5 & 6 & 7 & 8 & 9 & 10 \\
  \midrule
  \endhead

  \midrule
  \multicolumn{14}{r}{\textit{Continued on next page}} \\
  \endfoot

  \bottomrule
  \endlastfoot

  Houston Federation of Teachers Local 2415 v. Houston Independent School District (EVAAS) \ref{sec:houston}
  & Fourteenth Amendment procedural due process, substantive due process, Equal Protection
  & Score model and complete historical test score data for all students used in the calculation, necessary to independently verify and replicate an individual teacher's output score. All students and teachers' scores would have been necessary to replicate percentile score ranking.
  & Partial source code produced to plaintiff's expert under protective order, expert unable to replicate scores independently even with greater access than individual teachers
  & \cmark & \cmark & \cmark & \cmark & \xmark & \xmark & \xmark & \xmark & \xmark & \cmark \\
  \midrule

  TrueAllele cases \ref{sec:trueallele}
  & Admissibility of novel scientific evidence under varying evidentiary standards (including Frye, Daubert, and state-specific standards); Sixth Amendment right to confrontation
  & In \emph{Chubbs}: TrueAllele source code. In \emph{Pickett}: source code and software dependencies for the version used in the case, software development and testing documentation including bug reports and change logs, and underlying data and records from validation studies.
  & In \emph{Chubbs}: methodological descriptions, published articles, manuals, and supervised demonstrations offered by Cybergenetics; source code withheld. In \emph{Pickett}: same materials initially, source code ordered produced under protective order on remand, with format and terms of access heavily contested.
  & \cmark & \cmark & \cmark & \cmark & \xmark & \cmark & \cmark & \xmark & \cmark & \cmark \\
  \midrule

  State v. Loomis (COMPAS)
  & Fourteenth Amendment procedural due process
  & Disclosure of how COMPAS weights the risk score factors and calculates the scores, and information about the comparison population used to generate the assessment
  & Prior to litigation, each criminal defendant already had access to their inputs used to generate the risk scores, their risk score assessment in the presentence investigation report (PSI), and Northpointe's publicly available Practitioner's Guide. The court ruled that this was sufficient to satisfy procedural due process, and ordered no additional disclosures of the COMPAS algorithm.
  & \cmark & \cmark & \xmark & \cmark & \xmark & \xmark & \cmark & \xmark & \xmark & \cmark \\
  \midrule

  K.W. v. Armstrong \ref{sec:kw} & Fourteenth Amendment procedural due process; Medicaid Act \S~1396a(a)(3); ADA \S~202 (\emph{Olmstead} integration mandate) & Budget tool training data, variable weights, and constant coefficient; SIB-R assessment instrument used as key input to the tool; per-participant input and output data; accuracy and bias testing records & Full access ordered and obtained. Tool methods, training data, variable weights, constant coefficient, and per-participant input and output data all produced. SIB-R instrument produced under protective order. Accuracy and bias testing records sought but confirmed nonexistent: IDHW had conducted no such testing. & \cmark & \xmark & \xmark & \cmark & \xmark & \xmark & \xmark & \xmark & \xmark & \xmark \\
  \midrule

  MiDAS cases \ref{sec:midas} & \S 1983 Fourteenth Amendment procedural due process (\emph{Zynda}, \emph{Cahoo}); Fourth Amendment unreasonable seizure (\emph{Cahoo}); Michigan constitutional tort for money damages (\emph{Bauserman}) & Individual claimant unemployment compensation files; roster of all $\sim$44,000 affected claimants; files in underlying MiDAS databases; emails and hard drives of contractor employees assigned state accounts; SharePoint project files; documentation of customizations to MiDAS decision trees; internal agency communications about system performance; correspondence between the Unemployment Insurance Agency (UIA) and its contractors & Individual claimant files and 44,000 claimant roster produced under court order. Database files, decision tree documentation, and contractor emails ordered produced after eight months of non-compliance and no privilege log. Only 11 emails produced voluntarily. Multiple contractor employee mailboxes found deleted (some pre-litigation, some post-litigation hold). Deposition of UIA director permitted but scope-limited. Plaintiffs' independent expert precluded on procedural grounds. & \xmark & \xmark & \cmark & \cmark & \cmark & \cmark & \cmark & \cmark & \xmark & \cmark \\
  \midrule
  
  Connecticut Fair Housing Ctr v. CoreLogic Rental Property Solutions \ref{sec:cfhc} & Disparate impact and disparate treatment under FHA \S~3604(a); disability discrimination and failure to accommodate under FHA \S~3604(f); FCRA file disclosure claims; CUTPA & CrimSAFE proprietary matching algorithm, classification criteria, and source code; demographic output data by race and ethnicity across the full applicant pool; industry-wide configuration patterns across CoreLogic's Connecticut customer base; internal policy development records; plaintiffs' consumer screening report & Configuration matrix, lookback period settings, training materials, aggregate flagging rate data, WinnResidential's specific CrimSAFE configuration and access settings, and plaintiffs' complete screening report and adverse action letter produced in discovery. No algorithmic source code or matching logic produced. No demographic breakdown of flagged applicants by race or ethnicity obtained. & \xmark & \xmark & \cmark & \cmark & \xmark & \xmark & \xmark & \xmark & \xmark & \cmark \\
  \midrule

  Huskey v. State Farm Fire \& Casualty Co. \ref{sec:huskey} & Disparate impact under FHA \S~3604(b). & Tool identification and operational documentation for nearly two dozen algorithmic claims-routing and fraud-detection tools; development history, training data sourcing, and bias testing records; vendor contracts and third-party tool documentation; personnel training materials showing how claim handlers are instructed to interact with tool outputs; class-wide claims data including claimant demographics, processing timelines, documentation requests, and amounts paid; per-claim input and output data for each tool. & In discovery. Phase I complete; Phase II proceeding under June 2025 court order. Phase I produced approximately 14,700 pages and confirmed existence of nearly two dozen algorithmic tools through four 30(b)(6) depositions. Witnesses blocked from answering implementation-level questions as beyond the scope of Phase I. Vendor contracts and training materials beyond those specific to named tools denied. No claims-level data produced. Bias testing records unanswered. State Farm moved to proceed directly to class certification on the Phase I record before plaintiffs could obtain Phase II discovery; the court rejected this position and ordered discovery to proceed. & \cmark & \cmark & \cmark & \xmark & \xmark & \cmark & \cmark & \xmark & \xmark & \cmark \\
  \midrule
  
  Anderson v. TikTok \ref{sec:anderson} & Pennsylvania strict products liability (design defect) and negligence; wrongful death & Pre-discovery; no formal requests made. To support its claims, plaintiff would need evidence establishing that TikTok's personalized recommendation model specifically targeted Nylah Anderson with Blackout Challenge content. & Case dismissed on Section 230 grounds at pleading stage (E.D. Pa. 2022); reversed and remanded by the Third Circuit (Aug. 2024); TikTok declined to seek certiorari; case on remand, in pre-discovery stage. & \cmark & \cmark & \cmark & \xmark & \xmark & \xmark & \xmark & \xmark & \xmark & \cmark \\
  \midrule

  Mobley v. Workday (CSM) \ref{sec:mobley} & Disparate impact under Title VII, ADEA, and ADA, on an agent theory of liability. Disparate impact claims are proceeding toward Rule 23 class certification; the ADEA age discrimination claim has been separately certified as a collective action under the FLSA procedure (May 2025). & Evaluation data used to assess bias and demographic disparities in the model; internal audit documents and underlying impact data; applicant flow data across the employer-customers whose hiring decisions are at issue; technical documentation of how Workday's screening tools were developed and used. & In discovery. Court accepted Workday's representation that no evaluation data exists. Internal audit documents withheld under attorney-client privilege and work product doctrine; supplemental briefing ordered. Applicant flow data withheld on the basis that it is in employer-customers' possession, not Workday's, despite being stored on Workday's servers. Workday has agreed to produce training data for the Skills Embedding Model and documents describing CSM's inputs and functionality. & \cmark & \cmark & \cmark & \cmark & \cmark & \xmark & \cmark & \cmark & \xmark & \cmark \\
  \midrule
  
  Kadrey v. Meta \ref{sec:kadrey} & Direct copyright infringement under 17 U.S.C.\ \S~106 based on reproduction and distribution of plaintiffs' works in training Llama 1, 2, and 3 on shadow library datasets (LibGen, Anna's Archive). DMCA and CDAFA claims dismissed. & Composition and data mix proportions of training datasets across Llama versions; torrenting logs and records of shadow library acquisition; internal approvals and communications regarding data sourcing decisions; fine-tuning datasets including post-training safety data. & Training dataset composition and data mixes not produced. Post-training fine-tuning datasets for the ``Intellectual Property'' safety category ordered produced; raw source data underlying fine-tuning datasets denied on proportionality grounds. Guillaume Lample's 2022 LibGen torrenting logs produced April 2025 after corporate witnesses denied their existence. ML Hub database and AI Dataset Catalog dashboards, which showed internal approvals of shadow library acquisition, not searched during discovery; only discovered via hyperlink in a produced document. Court ordered search of ML Hub, Lample files, and additional AWS directories in November 2025. & \xmark & \xmark & \xmark & \cmark & \cmark & \xmark & \cmark & \cmark & \xmark & \xmark \\
  \midrule

  The New York Times Co. v. Microsoft Corp. \ref{sec:nyt} & Direct copyright infringement under 17 U.S.C.\ \S~106 based on use of plaintiffs' works to train GPT models and reproduction of those works in ChatGPT outputs. Consolidated with Daily News and other publisher actions; centralized into MDL in April 2025. & Training dataset composition and identity of works ingested across GPT model versions; Bing search index data; output log data showing reproduction of plaintiffs' content. & Training data composition withheld on commercial grounds; inspection offered only via controlled sandbox rather than direct production. Sandbox affected by technical disruptions; on November 14, 2024, OpenAI engineers erased plaintiffs' programs and search results, irretrievably losing folder structure and file names. Bing search index no longer available due to constant updating. Output log data not preserved since complaint filing; preservation order issued May 2025, upheld June 2025; terminated September 2025 and narrowed to domain-linked accounts. 20 million anonymized ChatGPT output logs ordered produced December 2025. & \xmark & \xmark & \xmark & \cmark & \cmark & \cmark & \cmark & \cmark & \xmark & \xmark \\
  \midrule

   Estate of Gene B. Lokken v. UnitedHealth Group, Inc \ref{sec:estate} & Breach of contract; breach of implied covenant of good faith and fair dealing. Plaintiffs claimed UHC breached Evidence of Coverage terms promising decisions by “clinical services staff” and “physicians” when it allegedly used AI. & Individual patient identities, "documents and communications about the development, design, creation, approval, implementation, and use of nH Predict," including "how nH Predict works, its development goals and anticipated benefit, and whether it was designed to supplant physician decision-making." Additionally, system level information such as "data, rules, source code, and medical guidelines."  &  "[D]ata, rules, source code, and medical guidelines" were denied. Everything else granted. & \xmark & \xmark & \cmark & \xmark & \xmark & \cmark & \xmark & \xmark & \xmark & \cmark \\
  \midrule   \\

    Chinook cases \ref{sec:chinook} & Judicial review under the \emph{Immigration and Refugee Protection Act}: breach of procedural fairness and fettering of statutory discretion through use of Chinook 3+ in temporary resident visa refusals. & \emph{Mehrara}: Chinook spreadsheets showing how applicant information was structured and presented to the decision-maker; ITAT risk indicator scores; officer working notes. \emph{Jahanian}: Same, plus note-generation interface logic governing pre-set refusal language. & \emph{Mehrara}: Spreadsheets deleted daily as routine IRCC practice and ruled unnecessary for production. ITAT risk score ruled irrelevant in \emph{Mehrara} because no score was assigned to the applicant. Working notes ruled marginally relevant only. \emph{Jahanian}: interface logic governing pre-set refusal language not produced; court remitted the decision for redetermination on the specific application without resolving the interface question. & \cmark & \xmark & \cmark & \xmark & \xmark & \xmark & \xmark & \cmark & \xmark & \xmark

\end{longtable}
}
}

\subsection{Case Analysis}
\label{sec:case_analysis}
\subsubsection{Houston Federation of Teachers Local 2415 v. Houston Independent School District (EVAAS)}
\label{sec:houston}
In Houston Federation of Teachers v. Houston Independent School District,\footnote{Houston Fed'n of Tchrs., Loc. 2415 v. Houston Indep. Sch. Dist., 251 F. Supp. 3d 1168 (S.D. Tex. 2017).} the Houston Independent School District (HISD) licensed EVAAS, a proprietary value-added model developed by SAS Institute, to make termination, nonrenewal, and bonus decisions for teachers based on student test score growth. Teachers and their union sued under the Fourteenth Amendment, alleging procedural due process, substantive due process, and equal protection violations. The court denied summary judgment on procedural due process, finding teachers had no meaningful way to verify their scores, but granted summary judgment for HISD on substantive due process and equal protection under rational basis review. The case settled with HISD agreeing not to use EVAAS as a basis to terminate teachers so long as those scores remain unverifiable.\footnote{Am. Fed'n of Teachers, \emph{Federal Suit Settlement: End of Value-Added Measures for Teacher Termination in Houston} (Oct. 10, 2017), AFT, \url{https://www.aft.org/press-release/federal-suit-settlement-end-value-added-measures-teacher-termination-houston}; \emph{see also} \emph{Houston Fed'n of Tchrs., Loc. 2415}, 251 F. Supp. 3d at 1175.}

Prior to litigation, teachers had access only to an overview of value-added methodology, a general description of EVAAS test methods, score report reading instructions, and current-year test scores for students taught by that teacher. HISD's claim that teachers had access to fuller student test score data was found unsupported by the record.\footnote{\emph{Houston Fed'n of Tchrs., Loc. 2415}, 251 F. Supp. 3d at 1183 n.38.} HISD argued at summary judgment that these materials satisfied due process, drawing an analogy to the general lab testing methods described in \emph{Banks v. FAA},\footnote{Banks v. Federal Aviation Admin., 687 F.2d 92 (5th Cir. 1982).} but the court rejected this, holding that describing EVAAS methodology was insufficient without the ability to verify the specific calculation (code 1).\footnote{\emph{Houston Fed'n of Tchrs., Loc. 2415}, 251 F. Supp. 3d at 1178-79.} For substantive due process, the court credited general academic community endorsement of value-added models despite HISD's admission that it had never independently verified or audited the scores, treating developer-adjacent validation as adequate under a permissive rational basis standard (code 2).\footnote{\emph{Id.} at 1181.} HISD also argued that EVAAS was merely one input into a holistic evaluation process with principals retaining final discretion, but the court found this claim was undermined by HISD's board goal of exiting 85\% of teachers with ineffective ratings and its evaluation of principals on whether they met teacher-exiting targets (code 3).\footnote{\emph{Id.} at 1180.}

SAS Institute treated the EVAAS source code as proprietary trade secrets, refusing to disclose them to either the plaintiff teachers or HISD itself. HISD consistently denied discovery requests for this information on the grounds that the source code was ``proprietary, trade secret information not in the custody, control, or possession of the District" (code 4).\footnote{\emph{Id.} at 1177 n.28.} During litigation, plaintiffs' expert received ``far greater access to the underlying computer codes than is available to an individual teacher" under a protective order, but was still unable to replicate the scores.\footnote{\emph{Id.} at 1177} Like in many risk assessment models, EVAAS estimated scores jointly across all teachers and students simultaneously and normed the scores against a statewide average that depended on every other teacher's data/scores. Thus, replicating a single teacher's predicted score would require access to the teacher's students' historical test score data and the scores of other teachers across the district and state. As HISD's own documentation acknowledged, correcting a single teacher's score would therefore require re-running the entire district-wide analysis because ``this re-analysis has the potential to change all other teachers' reports."\footnote{\emph{Id.} at 1177-78.} The court described this as a ``house-of-cards fragility" in which ``the accuracy of one score hinges upon the accuracy of all," meaning no requesting party could obtain the population-level access necessary to independently evaluate the accuracy of an individual score (code 10).\footnote{\emph{Id.} at 1178.} The court declined to order disclosure of SAS's proprietary information, reasoning that ``when a public agency adopts a policy of making high stakes employment decisions based on secret algorithms incompatible with minimum due process, the proper remedy is to overturn the policy, while leaving the trade secrets intact."\footnote{\emph{Id.} at 1179.} The parties ultimately settled on that basis, with HISD agreeing not to use unverifiable scores to terminate teachers.

\subsubsection{TrueAllele cases}
\label{sec:trueallele}
TrueAllele is a probabilistic genotyping software developed by Cybergenetics that uses a Markov chain Monte Carlo algorithm to analyze complex DNA mixtures and generate a likelihood ratio expressing the probability that a particular individual's DNA is present in a sample. Defendants in criminal proceedings repeatedly sought access to TrueAllele's source code to challenge admissibility of scientific evidence in pre-trial admissibility hearings under varying evidentiary standards — including Frye, Daubert, and state-specific standards — and to vindicate Sixth Amendment confrontation rights, resulting in a series of cases spanning over fifteen years across state and federal courts. We provide a full accounting of outcomes across evidentiary standards in Appendix \ref{sec:trueallele_all}; here we note that across different state and federal standards, courts have overwhelmingly declined to compel source code disclosure, a pattern that holds across jurisdictions and evidentiary standards, suggesting that privatization of proof in this context is largely standard-invariant. Here, we focus on two representative cases that reached opposite conclusions. In \emph{Chubbs}, the California Court of Appeal vacated a trial court order compelling source code disclosure, finding the defense had not made a sufficient particularized showing of need under Kelly/Frye.\footnote{People v. Superior Court (Chubbs), No. B258569, 2015 WL 139069 (Cal. Ct. App. Jan. 9, 2015).}. In \emph{Pickett}, the New Jersey Appellate Division ordered production under a protective order, finding independent adversarial review a prerequisite to a meaningful Frye reliability determination\footnote{State v. Pickett, 466 N.J. Super. 270 (App. Div. 2021).} \emph{Wakefield}\footnote{People v. Wakefield, 38 N.Y.3d 367 (2022).} is discussed for its treatment of the human-in-the-loop question implicated in Sixth Amendment right to confrontation claims.

Across the cases, Cybergenetics and the prosecution consistently argued that methodological descriptions, published articles, operating manuals, and over thirty validation studies , together with admissibility rulings from eighteen other jurisdictions, were sufficient to establish reliability without source code access (code 1).\footnote{Chubbs, 2015 WL 139069, at *2-3, *8; Pickett, 466 N.J. Super. at 307-08.} The \emph{Pickett} court rejected this, finding that none of this evidence addressed whether TrueAllele's source code correctly implemented its methods, and that the body of prior rulings constituted an authority ``house of cards", each having relied uncritically on the same developer-generated evidence (code 2).\footnote{\emph{Pickett}, 466 N.J. Super. at 308, 315-16.} The court catalogued that 35 of 36 validation studies were conducted by Cybergenetics or law enforcement, and 6 of 7 peer-reviewed publications were authored by Perlin himself, making the circularity of the lower court's crediting of these non-independent validation studies particularly acute in this case.\footnote{\emph{Id.} at 312-13.} In \emph{Wakefield}, the majority additionally held that meaningful human involvement in TrueAllele's workflow, the analyst's role in setting parameters and reviewing results, meant that cross-examining Dr. Perlin satisfied confrontation rights without source code access.\footnote{\emph{Wakefield}, 38 N.Y.3d at 385-86.} The concurrence rejected this directly, noting that Dr. Perlin had himself testified that ``TrueAllele does the work, we don't," suggesting that there was not meaningful human discretion over the algorithmic decisions, as argued by the producing party (code 3).\footnote{\emph{Id.} at 398-99 (Rivera, J., concurring).} On the production side, Cybergenetics invoked trade secrecy throughout, and in \emph{Pickett} offered access only under an NDA requiring inspection at the prosecutor's office under video surveillance, handwritten notes only, no electronic devices or internet access, a \$1 million automatic liability fine, and \$3 million insurance coverage, terms the court found so burdensome as to render meaningful review impossible (codes 4, 6, 7).\footnote{\emph{Pickett}, 466 N.J. Super. at 286-89, 317, 319-22.}

The TrueAllele cases raise distinctive questions about what access is necessary to evaluate a probabilistic prediction system. While validation studies can assess general system performance, the \emph{Pickett} court emphasized that they cannot substitute for source code review in detecting implementation-level errors — pointing to FST, a competing probabilistic genotyping software developer whose STRmix algorithm's source code was disclosed pursuant to a federal Daubert hearing, where an independent audit revealed a silent error excluding certain loci from the likelihood ratio calculation that had not been discovered through any prior validation study.\footnote{\emph{Pickett}, 466 N.J. Super. at 308, 313-14.} The version of TrueAllele used in \emph{Pickett} postdated every validation study cited by the State, meaning prior performance evaluations could not speak to the reliability of the specific implementation at issue (code 10).\footnote{\emph{Id.} at 313-14.} The court also identified an evaluability problem inherent to probabilistic predictions. While a system can be evaluated for calibration in aggregate, whether reported likelihoods are accurate across many cases, no individual likelihood ratio can be verified against a ground truth. For a given criminal defendant, there is only one sample and one prediction, and whether the modeling assumptions driving that particular output were appropriate cannot be determined from the output alone.\footnote{\emph{Id.} at 323.} Without source code access, a defendant cannot determine which specific implementation-specific choices are driving their likelihood ratio — how many contributors to assume, drop-in and drop-out parameter settings, which loci to include, how to handle peaks below the stochastic threshold (code 9) — whether they were appropriate, and whether alternative defensible choices would have produced a materially different result.\footnote{\emph{Id.} at 314.}

\subsubsection{State v. Loomis (COMPAS)}
\label{sec:loomis}
In \emph{State v. Loomis},\footnote{State v. Loomis, 371 Wis.2d 235 (Wis. 2016).} Eric Loomis pleaded guilty to two charges arising from a drive-by shooting in La Crosse, Wisconsin. His presentence investigation report (PSI) included a COMPAS risk assessment, which rated him as high risk on all three recidivism scales. The sentencing court referenced the COMPAS scores alongside other factors in ruling out probation and sentencing Loomis to six years of imprisonment. Loomis filed a post-conviction motion arguing that the circuit court's consideration of COMPAS violated his right to due process because its proprietary nature prevented him from challenging its scientific validity.

Loomis argued that due process required access to the score factors, weightings, and scoring methodology used to calculate his risk scores in COMPAS, and to the comparison population against which the assessment was made. The court held that each defendant's own 21 criminal history inputs listed in the PSI, the resulting risk scores, and Northpointe's publicly available Practitioner's Guide were sufficient to satisfy due process, crediting out-of-state validation studies as establishing the tool's reliability and ordering no disclosure of the algorithm, factor weightings, or comparison population (codes 1, 2).\footnote{\emph{Id.} at 259-61.} Northpointe treated COMPAS as a proprietary trade secret and was denied leave to appear as amicus curiae, yet its own publications and practitioner materials were extensively relied upon throughout the opinion as evidence of the tool's reliability (codes 2, 4).\footnote{\emph{Id.} at 286-87 (Abrahamson, J., concurring).}

The court's remedy is particularly notable for our purposes. Rather than treating the proprietary withholding of the algorithm as a barrier to its use, the court instead mandated that any PSI containing a COMPAS score include a written advisement stating that ``the proprietary nature of COMPAS has been invoked to prevent disclosure of information relating to how factors are weighed or how risk scores are determined.''\footnote{\emph{Id.} at 276.} The court thus explicitly acknowledges the opacity created by the trade secrecy claim as a limitation on the reliability of the assessment, yet having found procedural due process satisfied, its response is not to require disclosure but to ensure that sentencing courts are informed of that limitation (code 7). This is a direct illustration of the substitution dynamic we identify: structured disclosure about a system's limitations standing in for access to the system itself.

The case also illustrates two distinct performance evaluation problems (code 10). First, even with access to their own inputs and outputs, a defendant cannot determine whether their score reflects their individual risk profile or is instead driven by demographic characteristics, whether used directly or proxied through correlated inputs like neighborhood, employment history, or prior arrests, systematic biases of the model that might only be surfaced through black-box inference with counterfactual queries or access to outputs across a broader population of defendants. Second, the validity of any such evaluation depends on whether the reference population is appropriate: the court credited a New York study as establishing the tool's reliability while simultaneously acknowledging that ``no cross-validation study for a Wisconsin population has yet been completed.''\footnote{\emph{Id.} at 260.} Recidivism base rates differ across jurisdictions, and a finding that the tool predicts well for New York City probationers does not establish that it does so for Wisconsin defendants. The court required the PSI advisement to note this gap,\footnote{\emph{Id.} at 264, \P 66.} but credited the New York study as constitutionally sufficient to establish the tool's reliability.\footnote{\emph{Id.} at 260, \P\P 58-59.}

\subsubsection{K.W. v. Armstrong}
\label{sec:kw}
In \emph{K.W. v. Armstrong},\footnote{K.W. v. Armstrong, 180 F.Supp.3d 703 (D. Idaho 2016); \emph{see also} K.W. v. Armstrong, 789 F.3d 962 (9th Cir. 2015) (affirming preliminary injunction).} a certified class of developmentally disabled Medicaid recipients in Idaho sued the state Department of Health and Welfare after an automated budget calculation tool reduced their individualized service budgets without explanation. The tool used a statistical model with weighted variables and a constant coefficient to predict each participant's needs and set a corresponding budget, applying criteria that participants had no visibility into and no meaningful opportunity to contest. Plaintiffs brought claims under the Fourteenth Amendment procedural due process clause, the Medicaid Act, and the ADA's integration mandate under \emph{Olmstead}. Filed in 2012 and consolidated in 2013, the case was resolved at the district court on cross-motions for summary judgment in March 2016, following the Ninth Circuit's affirmance of a preliminary injunction in 2015. The court found the budget tool constitutionally deficient, ordered it redesigned with regular accuracy testing, and required IDHW to provide participants with written standards, adequate notice, representative assistance, and access to the materials used to calculate their budgets.

This case is primarily notable as a success story for evidentiary access, and both barriers that arose were overcome. The court rejected any argument that general descriptions sufficed and ordered full disclosure (code 1). Training data, variable weights, the constant coefficient, and per-participant input and output data were all produced, and the court ordered the tool itself to be redesigned and regularly tested going forward.\footnote{\emph{K.W.}, 180 F.Supp.3d at 724.} The one actively contested evidentiary access barrier was a copyright claim by Houghton Mifflin Harcourt, the publisher of the Scales of Independent Behavior --- Revised (SIB-R), a copyrighted assessment instrument used as a key input to the budget calculation. IDHW had restricted participants from using the SIB-R in appeal hearings on the publisher's objection. The court rejected this (code 4), holding that the risk of erroneous deprivation from excluding the instrument outweighed the publisher's commercial harm, and ordered a plan allowing participants to view and present the relevant portions of the SIB-R in appeals under a protective order.\footnote{\emph{Id.}\ at 717--18.} The accuracy and bias testing records that plaintiffs sought were found to be nonexistent: IDHW had never conducted such testing, and had been legally advised to halt any testing or improvement work during the litigation.\footnote{\emph{Id.}\ at 718.} This case differs from others in the case table in that the tool was government-built, relatively simple in architecture, and produced nearly complete access once the court found a due process violation. The SIB-R copyright barrier is nonetheless notable as an early instance of third-party intellectual property claims arising in the context of algorithmic accountability litigation, and the court's resolution of it in plaintiffs' favor offers a useful counterpoint to cases where such claims have succeeded in narrowing access.

\subsubsection{MiDAS cases}
\label{sec:midas}
Michigan's Unemployment Insurance Agency deployed MiDAS (Michigan Integrated Data Automated System) in October 2013 to automate fraud detection across unemployment compensation claims. MiDAS cross-checked claimant records against employer filings and federal databases, flagged discrepancies, and transmitted multiple-choice questionnaires to suspected claimants through online accounts. From October 2013 to August 2015, no human being took part in the adjudication process.\footnote{Cahoo v. SAS Analytics Inc., 912 F.3d 887, 893 (6th Cir. 2019).} If a claimant failed to respond within ten days or selected a triggering answer, MiDAS automatically issued a fraud determination, assessed penalties of up to four times benefits received, and initiated wage garnishment and tax refund interception without a pre-deprivation hearing. The Michigan Auditor General subsequently reviewed over 22,000 fraud determinations and found that 93\% did not involve actual fraud. Three litigations followed\footnote{\emph{Cahoo}, 912 F.3d at 894.}. \emph{Zynda v. Arwood} was a federal \S~1983 action against state officials challenging the system's due process and Fourth Amendment deficiencies, which settled in 2017, with the state voluntarily agreeing to suspend automated fraud determinations and conduct individualized review of prior cases as part of settlement negotiations\footnote{Zynda v. Arwood, 175 F. Supp. 3d 791 (E.D. Mich. 2016); 
Zynda v. Arwood, No. 2:15-cv-11449 (E.D. Mich. Feb. 2, 2017).}. As in \emph{Houston}, where HISD agreed not to use unverifiable EVAAS scores to terminate teachers, once the parties reached a settlement agreement, the litigation ended, meaning full evidentiary access to the system's internals was never obtained even where the underlying problems were serious enough to warrant restricting the system's use.\footnote{Am. Fed'n of Teachers, \emph{Federal Suit Settlement: End of Value-Added Measures for Teacher Termination in Houston} (Oct. 10, 2017), \url{https://www.aft.org/press-release/federal-suit-settlement-end-value-added-measures-teacher-termination-houston}; Houston Fed'n of Tchrs., Loc. 2415 v. Houston Indep. Sch. Dist., 251 F. Supp. 3d 1168, 1175 (S.D. Tex. 2017).} \emph{Bauserman v. Unemployment Insurance Agency} was a state court class action under the Michigan Constitution, which after two Michigan Supreme Court decisions resolving timeliness and the availability of money damages settled in 2024 for \$20 million covering approximately 3,000 claimants\footnote{Bauserman v. Unemployment Insurance Agency, No. 15-000202-MM (Mich. Ct. Claims May 10, 2016), \textit{aff'd in relevant part}, 509 Mich. 673 (2022).}. \emph{Cahoo v. SAS Analytics} was a federal \S~1983 suit targeting the private contractors who designed and implemented MiDAS, which produced the most extensive discovery record of the three and settled in 2023 for \$180,000 for four individual plaintiffs.\footnote{Cahoo v. SAS Analytics Inc., No. 2:17-cv-10657 (E.D. Mich. Jan. 17, 2024).}

The \emph{Cahoo} discovery record illustrates the full range of evidentiary access barriers. Plaintiffs sought implementation-level documentation of MiDAS's logic, including all customizations to the system's non-monetary determination decision trees and the rationale and authorship of each change. \footnote{Opinion and Order Granting Defendant FAST's Motion to Compel Compliance with Subpoena at 2, \emph{Cahoo}, No. 2:17-cv-10657 (E.D. Mich. Aug. 30, 2019), ECF No. 217.} The private contractors (Fast Enterprises, CSG, and SAS) asserted that MiDAS's decision trees, business rules, and 
implementation documentation constituted trade secrets and 
confidential commercial information, resisting disclosure on 
that basis (code 4).\footnote{Motion for Leave to File Items Under Seal at 3, \emph{Cahoo}, No. 2:17-cv-10657 (E.D. Mich. Sept. 8, 2020), ECF No. 469.} The Unemployment Insurance Agency separately invoked Michigan's unemployment compensation confidentiality statute, Mich.\ Comp.\ Laws \S~421.11(b)(1), and a parallel federal regulation to block production of both individual claimant files and a roster of the approximately 44,000 affected claimants, raising third-party privacy interests as a barrier to the core evidentiary record of the system's operation (code 5).\footnote{Opinion and Order Denying Motion to Quash Subpoena, \emph{Cahoo}, No. 2:17-cv-10657 (E.D. Mich. Apr. 23, 2019), ECF No. 186; Opinion and Order Denying Second Motion to Quash Subpoena, \emph{Cahoo}, No. 2:17-cv-10657 (E.D. Mich. June 25, 2019), ECF No. 210.} The court rejected both arguments, ordering production in each instance. 

Cost arguments were advanced by the state's Department of Technology, which estimated compliance costs potentially exceeding one million dollars, and by the Agency, which estimated \$1.5 million for privilege review of contractor emails alone; the court rejected these as grounds for withholding production, noting that privilege review costs are not reimbursable under Rule 45 because the producing party exclusively controls those costs (code 6).\footnote{Opinion and Order Granting Def. FAST's Mot. to Compel  Compliance with Subpoena at 7, \emph{Cahoo}, No.\ 2:17-cv-10657  (E.D. Mich. Aug. 30, 2019), ECF No. 217.} Negotiations over the format of electronically stored information production stalled for months. The Agency proposed producing claimant records by navigating the MiDAS interface screen by screen, printing each page, and scanning the output to PDF, a method whose cost and labor burden it had estimated in its own affidavit, while defendants' counsel argued that the same data could be extracted from the underlying database with a single query and exported directly to an electronic file, eliminating both the manual steps and the need for downstream OCR. The court declined to resolve the technical dispute without IT input from both sides, noting it could not determine whether proprietary constraints made native extraction genuinely impossible, but made clear that the producing party could not leverage its exclusive control over the data architecture to define the baseline production method in the most burdensome available terms.\footnote{Motion to Quash Hr'g Tr. at 21--22, \emph{Cahoo}, No. 2:17-cv-10657 (E.D. Mich. Feb. 6, 2019), ECF No. 208.} The court ultimately held that the entity controlling the data architecture bore the primary obligation to propose custodians and search terms, and that superior knowledge of one's own data could not be deployed to obstruct good-faith discovery.\footnote{Opinion and Order Denying Second Mot. to Quash  Subpoena at 10--11, \emph{Cahoo}, No.\ 2:17-cv-10657  (E.D. Mich. June 25, 2019), ECF No. 210.}
A similar asymmetry recurs in other algorithmic disclosure disputes, including \textit{New York Times Co. v. Microsoft Corp.}, where access to training data was mediated through a sandbox whose scope and query architecture were controlled entirely by the defendants.\footnote{See The New York Times Co. v. Microsoft Corp. (In re OpenAI Copyright Litig. MDL), No. 1:23-cv-11195 (S.D.N.Y. filed Dec. 27, 2023).} 

Despite a litigation hold issued when \emph{Bauserman} commenced, the Agency later disclosed that mailboxes for multiple named contractor employees were unavailable --- some deleted prior to any litigation, others after the hold was already in place --- implicating contested obligations around the preservation of system-adjacent artifacts and communications that bear directly on how the system operated and was overseen (code 8).\footnote{Opinion and Order Granting Def. FAST's Mot. to Compel  Compliance with Subpoena at 8, \emph{Cahoo}, No.\ 2:17-cv-10657  (E.D. Mich. Aug. 30, 2019), ECF No. 217.} The court ordered a targeted deposition on the circumstances, timing, and recoverability of the deletions. 

The individual agency defendants also argued, both at the motion to dismiss stage and on appeal, that MiDAS rather than any human decision-maker was responsible for the challenged deprivations, seeking to narrow the scope of their own liability and implicitly the scope of discoverable materials (code 3). The Sixth Circuit rejected this framing directly, finding that the defendants had implemented and oversaw MiDAS and had knowingly enforced its invalid determinations, and warning that accepting a qualified immunity defense grounded in technological novelty would give state actors a roadmap for constitutional evasion.\footnote{\emph{Cahoo}, 912 F.3d at 904--05.} 

Finally, plaintiffs' algorithmic accountability expert, Christian Sandvig, was precluded from testifying under Federal Rule of Civil Procedure 37(c)(1) for failure to comply with the disclosure requirements of Rule 26(a)(2)(B).\footnote{Opinion and Order Granting in Part and Denying in Part Motions to Strike Expert Witness Disclosures and Bar Expert Witness Testimony at 20--23, \emph{Cahoo}, No. 2:17-cv-10657 (E.D. Mich. Mar. 18, 2021), ECF No. 533.} The court found that Sandvig had been designated as a rebuttal witness, but that his proposed testimony --- that MiDAS failed to incorporate impartial decision-making, contained inherent biases, and produced predetermined outcomes --- went to elements of plaintiffs' case-in-chief rather than rebutting any specific opinion offered by defense witnesses. The defendants had strategically designated their witnesses primarily as fact witnesses rather than opinion experts, which meant there was little for a rebuttal expert to rebut, and the court held that the non-disclosure of a proper expert report was neither substantially justified nor harmless. Unlike in code 2, where the producing party affirmatively relies on developer-generated validation to foreclose independent scrutiny, here, the preclusion of the plaintiff's independent expert resulted from the interaction between defendants' strategic minimization of opinion testimony and plaintiffs' procedural missteps in resubmitting Sandvig after previously representing they would not use him as an expert. The expert pool capable of independently evaluating these systems is thin, and the combination of procedural constraints and adversarial disclosure strategy can effectively foreclose independent evaluation even where the requesting party has identified a qualified expert.

The MiDAS cases also illustrate what population-level performance evaluation requires and who is positioned to conduct it (code 10). The only systematic assessment of MiDAS's error rate came from the Michigan Auditor General's retrospective review of 22,000 cases --- a population-level audit that no individual claimant could have conducted and that postdated the period during which the system operated without human involvement. The Sixth Circuit incorporated the 93\% error rate directly into its procedural due process analysis under \emph{Mathews v. Eldridge}, treating it as establishing the system's profound risk of erroneous deprivation.\footnote{Cahoo v. SAS Analytics Inc., 912 F.3d at 902 (citing Mathews v. Eldridge, 424 U.S. 319, 335 (1976)).} Individual claimants received no information about the basis for their particular fraud flag, no access to the decision logic applied to their case, and no means to test whether the system treated similarly situated claimants differently.

\subsubsection{Connecticut Fair Housing Center v. CoreLogic Rental Property Solutions}
\label{sec:cfhc}
In \emph{Connecticut Fair Housing Center v. CoreLogic Rental Property Solutions},\footnote{Conn.\ Fair Hous.\ Ctr.\ v.\ CoreLogic Rental Prop.\ Sols., LLC, 167 F.4th 605 (2d Cir. 2026).} plaintiffs Connecticut Fair Housing Center and Carmen Arroyo, individually and as conservator of her son Mikhail Arroyo, sued CoreLogic Rental Property Solutions under the Fair Housing Act (FHA), the Fair Credit Reporting Act (FCRA), and the Connecticut Unfair Trade Practices Act (CUTPA). CoreLogic's automated tenant screening tool, CrimSAFE, flagged Mikhail as having a disqualifying criminal record on a shoplifting charge from 2014, below the level of a misdemeanor, that had already been dropped. The applicant was unable to see the reason for this flagging because CoreLogic's report provided no details about his underlying criminal history to the landlord, only a computer-generated notation that the application did not meet the landlord's criteria. A separate dispute arose from CoreLogic's refusal to provide Carmen Arroyo a copy of Mikhail's consumer file: CoreLogic required a notarized power of attorney as a condition of third-party disclosure, a document that Mikhail, as a conserved person, lacked the legal capacity to execute. Filed in 2018, the case proceeded through cross-motions for summary judgment in 2020 and a ten-day bench trial concluded in November 2022. The district court found for CoreLogic on the FHA and CUTPA claims and for Mikhail Arroyo on a narrow FCRA theory, awarding statutory and punitive damages.\footnote{\emph{CFHC v.\ CoreLogic Rental Prop.\ Sols., LLC}, No.\ 18-CV-705, 2023 WL 4669482, at *23--25 (D.\ Conn.\ July 20, 2023).} The Second Circuit dismissed CFHC's claims for lack of organizational standing, affirmed the FHA judgment on proximate causation grounds, and reversed the FCRA award, leaving CoreLogic with no adverse judgment on any claim.\footnote{167 F.4th at 632.}

Discovery proceeded on the FHA claims, which survived summary judgment in 2020, and produced substantial documentation of CrimSAFE's operation, including its classification criteria, lookback period structure, training materials, aggregate flagging rate data, and WinnResidential's specific configuration settings, as well as plaintiffs' complete screening report and adverse action letter.\footnote{\emph{CFHC v.\ CoreLogic Rental Prop.\ Sols., LLC}, No.\ 18-CV-705, 2023 WL 4669482, at \P\P 7--8, 11, 21--22, 25--26, 30--31, 36 (D.\ Conn.\ July 20, 2023).} The matching logic and source code underlying CrimSAFE's categorization were not among the materials produced. Both courts resolved the question of whether CrimSAFE was a decision-maker as a threshold matter rather than proceeding to merits-level inquiry into the tool's implementation (code 3). That characterization is worth noting alongside CrimSAFE's suppression feature: WinnResidential had configured the tool to withhold the underlying criminal record from its on-site leasing staff, who saw only the binary ``Record(s) Found'' output and were structurally prevented from exercising independent judgment about the record.\footnote{\emph{CFHC}, No.\ 18-CV-705, 2023 WL 4669482, at \P 26.} The legal conclusion that WinnResidential retained discretion was thus in some tension with the operational reality that CrimSAFE's architecture constrained that discretion at the ground level, but that tension was never examined through access to the tool's implementation. A distinct evidentiary gap concerns population-level evaluation (code 10). CoreLogic tracked race only for applicants who matched with a criminal record, not across the full applicant pool, meaning the system was never designed in a way that would allow plaintiffs to obtain the demographic data necessary to evaluate CrimSAFE's disparate impact across its deployment population, a structural absence in CoreLogic's data collection practices that no production order could have remedied.

CoreLogic's failure to recognize conservatorship as valid identification, instead requiring a power of attorney that a conserved person cannot execute, is more of a flaw in CoreLogic's identification policy than an access issue specific to automated tooling.

\subsubsection{Huskey v. State Farm Fire \& Casualty Co.}
\label{sec:huskey}
In \emph{Huskey v. State Farm Fire \& Casualty Co.},\footnote{Huskey v. State Farm Fire \& Cas. Co., No. 22-CV-7014, 2023 WL 5848164 (N.D. Ill. Sept. 11, 2023).} Jacqueline Huskey and Riian Wynn, two Black homeowners insurance policyholders, brought disparate impact claims under \S~3604(b) of the Fair Housing Act, alleging that State Farm's use of algorithmic fraud-detection and claims-routing tools produced statistically significant racial disparities in claims processing times, documentation requirements, and volume of interactions required of Black claimants. A 2021 survey of approximately 800 State Farm policyholders found that white claimants were nearly one-third more likely to have their claims processed within a month, and that Black claimants were 39\% more likely to be required to submit additional documentation.\footnote{\emph{Id.} at *2.} The court denied State Farm's motion to dismiss in September 2023, finding the complaint plausibly alleged a statistical disparity, a specific policy, and a causal connection.\footnote{\emph{Id.} at *8--9.} Claims under \S~3604(a) and \S~3605 were dismissed. The case is in active discovery, with Phase I complete and Phase II proceeding under a June 2025 court order.

A central question in \emph{Huskey} is a variant of the dispute that animates class certification doctrine under \emph{Wal-Mart Stores, Inc. v. Dukes}: whether the algorithmic tools constitute a common, binding decision-making mechanism that unifies the class, or whether individual claim handler discretion is the operative variable such that claims turn on unique facts rather than a shared policy (code 3). State Farm has consistently advanced the latter position, asserting through corporate witnesses that ``no matter the FAE to which a claim is routed, the actual handling of that claim will depend upon factors unique to that claim and the discretion of the claim representative assigned to handle it.''\footnote{Joint Status Report Per January 14, 2025 and February 5, 2025 Orders at 21, \emph{Huskey}, No. 22-cv-7014 (N.D. Ill. May 15, 2025), ECF No. 166.} This framing simultaneously anchors State Farm's defense to class certification and its basis for limiting Phase I discovery to tool identification, blocking investigation into how the tools were developed, how their outputs flow into claim handling, and whether they were tested for racial bias. The deposition record illustrates how scope limitations operated in practice: corporate witnesses could not answer basic questions about tool goals, whether SAS tools had bias-detection capabilities, how claims move between systems step by step, or whether tools use predictive modeling, with State Farm's counsel repeatedly interposing ``beyond the scope'' objections throughout.\footnote{\emph{Id.} at 6--7.} These objections were grounded in the magistrate judge's Phase I order, which limited initial discovery to tool identification and explicitly reserved ``communications and documents about their use, and the means and methods in which they are used'' for a later phase, directing the parties to ``identify the tools first \ldots\ before the deep dive into the tools.''\footnote{Joint Status Report Per 2/23/24 Order at 1, \emph{Huskey}, No. 22-cv-7014 (N.D. Ill. Mar. 1, 2024), ECF No. 80 (quoting Dkt. 79).} Phase I produced no claims-level data whatsoever, and the question of whether any tool was evaluated for racial bias was never answered (codes 1, 2).

State Farm's litigation strategy exploited the phased discovery structure in two directions. On access, it successfully obtained denial of vendor contracts and training materials beyond those specific to named tools on proportionality and undue burden grounds (code 6). The training materials are particularly significant because without them plaintiffs cannot confirm or rebut State Farm's assertion that claim handlers exercise independent judgment rather than following tool outputs. Having restricted Phase I to tool identification, State Farm then moved to proceed directly to class certification on that thin record, arguing the evidence already showed no class could be certified, a position the court rejected in June 2025, ordering Phase II to proceed.\footnote{Minute Entry, \emph{Huskey}, No. 22-cv-7014 (N.D. Ill. June 10, 2025), ECF No. 173.} Plaintiffs' Phase II proposals make explicit what population-level disparate impact analysis requires: per-claim input and output data for each tool, class-wide claims data broken down by claimant race, processing timelines, documentation requests, and amounts paid (code 10). State Farm's response adds a further dimension to the access problem. Even if plaintiffs obtained class-wide claims data and demonstrated aggregate racial disparities in processing times, State Farm argues under \emph{Dukes} that this would not suffice to satisfy Rule 23's commonality requirement, because aggregate statistical evidence cannot demonstrate that the tools rather than individual claim handlers produced those disparities.\footnote{Joint Status Report Per January 14, 2025 and  February 5, 2025 Orders at 21--22, \emph{Huskey}, No.\ 22-cv-7014  (N.D. Ill. May 15, 2025), ECF No. 166.} The evidence that would actually answer that question is precisely what Phase II seeks: per-claim input and output data for each tool, which would allow plaintiffs to trace whether tool outputs correlate with disparate outcomes at the individual claim level rather than in the aggregate alone (code 7).

\subsubsection{Anderson v. TikTok}
\label{sec:anderson}
In \emph{Anderson v. TikTok},\footnote{Anderson v. TikTok Inc., 116 F.4th 180 (3d Cir. 2024).} Tawainna Anderson sued TikTok and its parent ByteDance following the death of her ten-year-old daughter Nylah, who died in December 2021 after attempting the ``Blackout Challenge'' --- a viral trend encouraging self-asphyxiation --- promoted via videos recommended to her by TikTok's For You Page (FYP) algorithm. Anderson brought claims under Pennsylvania strict products liability and negligence, alleging the FYP algorithm was defectively designed to surface dangerous content to vulnerable minors. The Eastern District of Pennsylvania dismissed the complaint on Section 230 grounds, finding TikTok immune as a publisher of third-party content. The Third Circuit reversed, holding that TikTok's algorithmic recommendations constitute the platform's own expressive activity --- first-party speech --- rather than the publication of user-generated content, and that Section 230 does not bar claims premised on the algorithm's affirmative promotion of content.\footnote{\emph{Id.} at 183--84.} TikTok declined to petition for certiorari, leaving the Third Circuit ruling as settled precedent in that circuit. The case was remanded to the Eastern District of Pennsylvania and remains pre-discovery.
\emph{Anderson} illustrates how threshold doctrinal disputes about algorithmic role can function as a structural evidentiary barrier, entirely antecedent to formal discovery. The district court's dismissal, and the pre-\emph{Anderson} approach by circuit courts in \emph{Force v. Facebook}\footnote{Force v. Facebook, Inc., 934 F.3d 53 (2d Cir. 2019).} and \emph{Gonzalez v. Google},\footnote{Gonzalez v. Google LLC, 2 F.4th 871 (9th Cir. 2021), vacated and remanded, 598 U.S. 617 (2023).} resolved the character of ML-based recommendation systems using a generic methodological description --- treating the FYP as a content-neutral sorter of third-party content functionally equivalent to a traditional publisher's editorial judgment, without any implementation-level inquiry into how the model actually ranks content for individual users. The \emph{Force} majority held that Facebook's algorithms operate on ``objective factors applicable to any content, whether it concerns soccer, Picasso, or plumbers''\footnote{\emph{Force}, 934 F.3d at 70.}, a characterization that treats personalized ML ranking as categorically equivalent to alphabetical sorting, eliding the entire implementation layer where consequential, user-specific decisions are made. As one commentator observed of \emph{Gonzalez}, the parties' only factual submission in the entire record of how targeted recommendation algorithms work was a single screenshot of the results of a YouTube search that plaintiffs reproduced in their complaint.\footnote{\emph{See} Adam Candeub, \emph{Gonzalez, Google, and Section 230: All on the Same Side?}, Fed. Soc'y Blog (Feb. 2, 2023).} The Section 230 immunity determination was thus made (across the circuit split, in both directions) on the basis of an abstract description of algorithmic function supplied by the parties, without independent technical review of how any platform's model actually targets individual users (codes 1, 2, 3). This is structurally similar to the dynamic \emph{Pickett} identified in TrueAllele, a legal determination about a system's function and reliability resolved on the basis of the producing party's characterization of the system, without independent implementation-level inquiry.

Proving causation on remand will require showing not merely that Blackout Challenge content appeared on Nylah's FYP, but that the model's personalized targeting drove her engagement with it in a way sufficient to establish products liability. That showing requires population-level inference into how the model behaves across networked users at scale, whether it systematically amplified this content category to users with particular demographic or behavioral profiles, and whether its targeting of vulnerable minors with dangerous content was a feature of the model's optimization (code 10). This barrier may be more difficult to overcome for claims premised on internal algorithmic features than for claims premised on external design features. The social media addiction litigation, for instance, has proceeded in part because plaintiffs could document visible product features like infinite scroll and autoplay through user testimony and internal company documents.\footnote{\emph{See} Plaintiffs' Second Amended Master Complaint (Personal Injury) \P\P~125--32; \emph{In re} Social Media Adolescent Addiction/Pers. Inj. Prods. Liab. Litig., No. 4:22-md-03047-YGR (N.D. Cal. Dec. 15, 2023), ECF No. 494.} Whether a recommendation model specifically targeted a particular user with particular content is harder to establish from outside the system, since evaluating that claim plausibly requires some understanding of what the model is optimizing for, facts that are internal to the system and not readily recoverable from its outputs alone.

\subsubsection{Mobley v. Workday}
\label{sec:mobley}
In \emph{Mobley v. Workday},\footnote{Mobley v. Workday, Inc., 740 F. Supp. 3d 796 (N.D. Cal. 2024).} Derek Mobley, an African American male over forty with disclosed anxiety and depression, brought employment discrimination claims against Workday, alleging that its AI-powered applicant screening tools systematically rejected him across more than one hundred job applications on the basis of race, age, and disability. Mobley brings disparate impact claims under Title VII, the ADEA, and the ADA, proceeding on an agent theory of liability: that Workday's customers delegated traditional hiring functions to Workday's algorithmic tools, making Workday liable as an employer's agent under the federal anti-discrimination statutes. The court denied Workday's motion to dismiss on the disparate impact claims in July 2024, finding that the speed and pattern of Mobley's rejections (including one received at 1:50 a.m., less than an hour after submission) gave rise to a plausible inference of automated screening on protected characteristics.\footnote{\emph{Id.} at 811.} The ADEA collective was conditionally certified in May 2025 under the FLSA procedure; Rule 23 class certification on the remaining claims is slated for 2026. The case is in active discovery.

Workday characterizes its Candidate Skills Match (CSM) tool as mechanically comparing skills extracted from job postings against skills extracted from resumes --- analogous, in its framing, to a spreadsheet or email system carrying out instructions rather than exercising independent judgment (code 3). The court rejected this characterization at the pleading stage, noting that Workday's tools are alleged to ``automatically disposition'' candidates and ``make hiring decisions'' through embedded AI and machine learning, not merely filter on minimum qualifications.\footnote{\emph{Id.} at 807.} The same definitional dispute has migrated into discovery. When plaintiffs sought evaluation data used to assess bias and demographic disparities (RFP 3), Workday responded that CSM is ``an algorithm, not a model'' and that therefore no evaluation data exists (code 1, code 10).\footnote{Parties' Joint Letter Regarding Discovery Dispute at 4, \emph{Mobley}, No. 3:23-cv-00770-RFL (N.D. Cal. May 8, 2025), ECF No. 124.} Plaintiffs countered that Workday's own AI Fact Sheet acknowledges it ``performed statistical testing to detect impermissible bias'' in CSM, directly contradicting the representation that no such data exists.\footnote{\emph{Id.} at 1--2.} The court accepted Workday's representation without further inquiry (code 2). The implications are significant: if a vendor's self-serving terminological characterization (algorithm versus model) is accepted as dispositive without independent review, the absence of evaluation data becomes self-insulating, creating a structural incentive to forgo bias evaluation altogether rather than generate discoverable evidence of disparate impact.

The internal audit documents present a related problem. Workday has asserted attorney-client privilege and work product protection over audits conducted at the direction of in-house counsel, arguing further that even the underlying data constitutes attorney work product because counsel curated which data fields to include in the analysis.\footnote{\emph{Id.} at 4--5.} This assertion that factual impact data is shielded because an attorney selected it sits at the boundary of ordinary work product, which can be overcome on a showing of substantial need and inability to obtain the equivalent without undue hardship, and opinion work product, which receives near-absolute protection. The court ordered supplemental briefing rather than resolving the dispute (code 4). Notably, Workday also represents that it ``no longer has access to even the limited, attorney work product dataset'' used in the privileged audit, raising a question about whether that data was preserved after the attorney-directed analysis concluded or has since been lost (code 8).\footnote{\emph{Id.} at 5.}

Workday separately argues that plaintiffs need not invade privilege because they can conduct their own disparate impact analysis using applicant datasets available for purchase from third-party vendors, or using the ``forthcoming training data set'' that Workday has agreed to produce, in effect proposing that plaintiffs obtain equivalent access through alternative means that Workday itself identifies and controls (code 7).\footnote{\emph{Id.} at 5.} The training data offer is likely narrower than it appears. Based on the discovery record, Workday's offer concerns the Skills Embedding Model training corpus, the data from which the model learned skills representations, rather than applicant flow data showing how actual candidates were scored and dispositioned. These are categorically different for purposes of a disparate impact analysis: training data illuminates what the model learned from, but applicant flow data (who applied, what scores they received, whether they advanced) is what is needed to calculate whether protected groups were disproportionately rejected. Workday's offer to substitute the former for the latter, if accepted, would leave plaintiffs with access to the inputs to model training but not the outputs of model deployment against real applicants needed for population-level disparate impact analysis (code 10).

Applicant flow data across employer customers has been withheld on a dual basis: Workday claims the data is in customers' possession and custody, while customers have responded to subpoenas by directing plaintiffs back to Workday, leaving plaintiffs caught between two parties each denying control over data stored on Workday's servers (code 5).\footnote{\emph{Id.} at 3--5.} This is a structural feature of cloud-based AI deployment that differs from earlier algorithmic accountability cases: because Workday operates as a data processor on behalf of employer-customers who retain formal data ownership, the standard discovery path to applicant-level data is blocked at both ends, even though the data physically resides on Workday's infrastructure and Workday processed it to produce the scores at issue. A further complication is that Workday's tools are customized per employer, meaning the feature weights and decision logic applied to Mobley's applications varied across the more than one hundred employers at issue. Establishing a common discriminatory mechanism across that variation requires implementation-level access to each employer's deployment configuration, not just a description of how CSM works in general (code 1).

\subsubsection{Kadrey v. Meta Platforms, Inc.}
\label{sec:kadrey}
\emph{Kadrey v. Meta Platforms, Inc.}\footnote{Kadrey v. Meta
Platforms, Inc., No.\ 3:23-cv-03417 (N.D. Cal. filed July 7, 2023).}
is a putative class action brought by thirteen authors against Meta,
alleging direct copyright infringement under 17 U.S.C.\ \S~106 based
on Meta's use of their works, obtained from shadow libraries including Library Genesis and Anna's Archive, to train its Llama large language models. Most claims were dismissed in November 2023, with direct copyright infringement surviving. Following consolidation with a related action and multiple rounds of amendment, the case proceeded through extensive fact and expert discovery. On June 25, 2025, Judge Chhabria granted Meta's cross-motion for summary judgment on the fair use question, holding that Meta's training use was highly transformative and that plaintiffs had not demonstrated market harm.\footnote{Order Denying Pls.' Mot. for Partial Summ. J.\ and Granting Meta's Cross-Mot. for Partial Summ. J.\ at 28, Kadrey v.\ Meta Platforms, Inc., No.\ 3:23-cv-03417 (N.D. Cal. June 25, 2025), ECF No.\ 598.} The DMCA and CDAFA claims were separately dismissed. A distribution claim and a fourth amended complaint proceed as of March 2026.\footnote{Order Granting Leave to File Fourth Am.\ Compl., Kadrey, No.\ 3:23-cv-03417 (N.D. Cal. Mar.\ 25, 2026), ECF No.\ 700.}

A notable evidentiary feature of \emph{Kadrey} is that the record on
which the fair use ruling rested was shaped at multiple levels by
training data withholding. Meta's own Llama~2 FAQ stated that data
mixes were ``intentionally withheld for competitive reasons,'' and
plaintiffs alleged this framing served to avoid scrutiny of copyright infringement rather than to protect genuine competitive
interests.\footnote{Corrected Second Consolidated Am.\ Compl.\
\P~72, at 12, Kadrey, No.\ 3:23-cv-03417 (N.D. Cal. Sept.\ 9,
2024), ECF No.\ 133.} Plaintiffs were left to establish infringement
through proxy methods, public dataset documentation and targeted
output queries, rather than direct inspection of which works were
incorporated across model versions (code~4).

The court denied production of the raw source data underlying Meta's
supervised fine-tuning datasets on proportionality grounds, relying on a passage from Meta's own Llama~2 technical paper describing that raw data as ``massive compared to the datasets actually used'' to conclude that production was not proportional to the needs of the
case.\footnote{Public Version of Discovery Order at ECF No.\ 374, at
4--5, Kadrey, No.\ 3:23-cv-03417 (N.D. Cal. Jan.\ 8, 2025), ECF
No.\ 399.} The passage in question was written by Meta to describe a
deliberate design choice: it had set aside millions of lower-quality
third-party supervised fine-tuning examples in favor of fewer,
higher-quality vendor-annotated examples, a decision it presented as a technical contribution. The court repurposed this self-characterization of a curation methodology as a finding about discovery burden. This is adjacent to the dynamic code~2 identifies: rather than crediting developer-generated evidence to establish the system's reliability, the court credited the producing party's own technical paper to establish the cost of producing it --- with the same effect of limiting access through reliance on characterizations that only the producing party was positioned to make. The court separately ordered production of post-training datasets for the ``Intellectual Property'' safety fine-tuning category, finding those relevant and proportionate (code~6).\footnote{\emph{Id.}}

The retention record reflects a recurring pattern of incomplete
disclosure corrected only through plaintiff investigation (code~8).
Torrenting logs from ex-employee Guillaume Lample documenting Meta's
Fall 2022 LibGen acquisition were produced only in April 2025, after
Meta's corporate designee had testified that the company had ``not
been able to track down any evidence of what had been downloaded or
been able to find that -- any more details.''\footnote{Pls.'
Administrative Mot.\ Requesting Leave to Obtain Limited Additional
Discovery at 2, Kadrey, No.\ 3:23-cv-03417 (N.D. Cal. Nov.\ 20,
2025), ECF No.\ 654.} Meta's centralized ML Hub database and AI
Dataset Catalog (AIDC) dashboards, which showed that Meta approved acquiring data from sites the federal government had designated notorious piracy markets, and that Meta's own Policy team had recommended against training on such sites, were never searched during discovery. Plaintiffs discovered them only because a produced document contained a hyperlink to them.\footnote{\emph{Id.}\ at 2--3.} An additional AWS S3 repository containing 2024 torrenting evidence was not searched until September 2025, and when searched revealed direct evidence of LibGen Fiction torrenting that Meta had previously represented did not occur.\footnote{\emph{Id.}\ at 3.} The court granted additional discovery into all three threads in November 2025.\footnote{Order Granting Pls.' Mot.\ for Additional Discovery, \emph{Kadrey}, No.\ 3:23-cv-03417 (N.D. Cal. Nov.\ 13, 2025), ECF No.\ 647.}

Judge Chhabria's observation that cases with better-developed
evidentiary records on market harm would likely come out
differently\footnote{Order Denying Pls.' Mot. for Partial Summ.\ J.\
and Granting Meta's Cross-Mot. for Partial Summ.\ J.\ at 28, \emph{Kadrey},
No.\ 3:23-cv-03417 (N.D. Cal. June 25, 2025), ECF No.\ 598.} takes
on additional significance given that the market harm record was
itself shaped by what the training data composition record did and did not contain. The case illustrates that competitive withholding of training data composition (code~4) and scale-based proportionality arguments (code~6) can jointly foreclose the evidentiary record on which the merits of the case ultimately turn, even where successive post-judgment disclosures suggest that record was materially
incomplete.

\subsubsection{The New York Times Co. v. Microsoft Corp.}
\label{sec:nyt}
\emph{The New York Times Company v. Microsoft Corporation}\footnote{The New York Times Co.\ v.\ Microsoft Corp., No.\ 1:23-cv-11195 (S.D.N.Y. filed Dec.\ 27, 2023).} is a copyright infringement action brought by The New York Times Company and Daily News, LP against Microsoft and OpenAI, alleging that defendants trained the GPT family of large language models on plaintiffs' copyrighted works without authorization and that ChatGPT outputs directly infringe those works by reproducing substantial portions of their content. The original \emph{Times} action was filed in December 2023; related actions by the Daily News and other newspaper publishers were consolidated with it in September 2024 and subsequently centralized into a broader MDL.\footnote{\emph{In re} OpenAI, Inc.\ Copyright Infringement Litig., No.\ 1:25-md-3143 (S.D.N.Y. centralized Apr.\ 3, 2025).} The case is in active discovery with no merits resolution.

OpenAI justified its withholding of training data composition on commercial grounds. 
(code~4).\footnote{Complaint \P~60, \emph{New York Times Co.}, No.\ 1:23-cv-11195 (S.D.N.Y. Dec.\ 27, 2023), ECF No.\ 1.} Rather than searching its own datasets, OpenAI responded to all of plaintiffs' discovery requests by offering inspection in ``a very, very tightly controlled environment'' that the court referred to as ``the sandbox'' --- externalizing the technical labor, error risk, and verification burden onto plaintiffs despite OpenAI's exclusive control over and familiarity with its own data (codes~6, 7).\footnote{Joint Letter re Training Data Issues at 1, \emph{New York Times Co.}, No.\ 1:23-cv-11195 (S.D.N.Y. Nov.\ 1, 2024), ECF No.\ 305 (quoting Sept.\ 12, 2024 Hearing Tr.\ at 5:23--6:6).} The parties ultimately agreed to a hybrid methodology combining six-word n-gram token matching with URL-based domain matching, though plaintiffs noted this approach would undercapture works that had been republished, paraphrased, or otherwise modified (code~7).\footnote{Letter re Training Data Issues at 1--2, \emph{New York Times Co.}, No.\ 1:23-cv-11195 (S.D.N.Y. Nov.\ 20, 2024), ECF No.\ 328.}

The retention record produced three distinct failures (code~8). First, on November 14, 2024, after plaintiffs had spent over 150 person-hours searching in the sandbox, OpenAI engineers erased all of plaintiffs' programs and search result data from one of the dedicated virtual machines; while OpenAI recovered much of the underlying data, the folder structure and file names were irretrievably lost, rendering the recovered data unreliable for tracing where plaintiffs' articles appeared in the training corpus.\footnote{Letter re Training Data Issues at 1, \emph{New York Times Co.}, No.\ 1:23-cv-11195 (S.D.N.Y. Nov.\ 20, 2024), ECF No.\ 328.} Second, Bing search index data previously transferred to OpenAI was no longer available due to constant updating. Third, it was not until November 15, 2024 --- nearly a year after the complaint was filed --- that plaintiffs learned OpenAI had been failing to preserve output log data since at least the date of the initial complaint, despite a litigation hold letter and explicit requests through multiple discovery vehicles.\footnote{Letter re Output Log Preservation at 1--2, \emph{New York Times Co.}, No.\ 1:23-cv-11195 (S.D.N.Y. Jan.\ 13, 2025), ECF No.\ 379.}

The output log dispute illustrates how privacy and retention concerns interact (codes~5, 8). OpenAI resisted preservation by invoking user privacy interests and ``numerous privacy laws and regulations throughout the country and the world,'' characterizing a broad preservation order as inconsistent with user expectations and contractual commitments.\footnote{Preservation Order at 2, \emph{In re} OpenAI, Inc.\ Copyright Infringement Litig., No.\ 1:25-md-3143 (S.D.N.Y. May 13, 2025), ECF No.\ 33.} OpenAI's CEO publicly proposed a novel ``AI privilege'' governing user conversations, an early instance of what may become a broader pattern of model developers invoking user privacy interests to categorically block disclosure of inference inputs and outputs.\footnote{\emph{See} Sarah Perez, \emph{Sam Altman Warns There's No Legal Confidentiality When Using ChatGPT as a Therapist}, TechCrunch (July 25, 2025), \url{https://techcrunch.com/2025/07/25/sam-altman-warns-theres-no-legal-confidentiality-when-using-chatgpt-as-a-therapist/}.} Following OpenAI's failure to confirm whether it would take any preservation steps absent a court order, Magistrate Judge Wang on May 13, 2025 directed OpenAI to preserve and segregate all output log data that would otherwise be deleted on a going-forward basis.\footnote{Order, \emph{New York Times Co.}, No.\ 1:23-cv-11195 (S.D.N.Y. May 13, 2025), ECF No.\ 33.} District Judge Stein denied OpenAI's objection in response to Magistrate Wang's order.\footnote{Order, \emph{New York Times Co.}, No.\ 1:23-cv-11195 (S.D.N.Y. June 26, 2025), ECF No.\ 712.} The parties subsequently agreed to a narrowed resolution terminating OpenAI's ongoing preservation obligation as of September 26, 2025, retaining data already preserved and requiring going-forward preservation of output logs associated with domains identified by the News Plaintiffs.\footnote{Stipulation and Order to Terminate OpenAI's Ongoing Obligations Under the Preservation Order, \emph{New York Times Co.}, No.\ 1:23-cv-11195 (S.D.N.Y. Oct.\ 9, 2025), ECF No.\ 922.}

\subsubsection{The Estate of Gene Lokken v. UnitedHealth Group}
\label{sec:estate}
In The Estate of Gene B. Lokken v. UnitedHealth Group, plaintiffs allege that UnitedHealth used nH Predict to systematically deny post-acute care to Medicare Advantage patients, breaching its contractual commitment to physician-led clinical review. Plaintiffs allege that the tool applied rigid criteria for post-acute care which effectively limited individualized physician judgment UnitedHealth's own policy documents promised. UnitedHealth counters that nH Predict is merely a “guide” and that physicians retain final decision-making authority. However, the system's 90\% error rate had horrifying effects for such a sensitive population; many elderly patients died before seeing a resolution or being able to contest their claims. Only .2\% of policyholders would file an appeal to their denials \footnote{\textit{The Estate of Gene B. Lokken et al. v. UnitedHealth Group, Inc. et al.}, No. 0:23-cv-03514 (D. Minn. Nov. 14, 2023).}. The case is proceeding on claims of breach of contract and breach of the implied covenant of good faith and fair dealing to proceed, highlighting the relevance of the ``role of the AI actor" in critical decision making scenarios.

The evidentiary contest in \emph{Lokken} has played out primarily through UnitedHealth's repeated efforts to confine discovery by moving in August 2025 to bifurcate discovery into two stages.\footnote{\emph{The Estate of Gene B. Lokken}, 2025 WL 2607196, at *1.} Under UnitedHealth's proposal, stage-one discovery would have been limited to ``whether Defendants used nH Predict in place of physician medical directors to make adverse coverage determinations related to the named Plaintiffs' care'', with expert and class-based discovery proceeding only if plaintiffs survived a stage-one summary judgment motion on that question.\footnote{\emph{Id.}} The proposal effectively renders \emph{whether} a physician formally signed off on a denial the dispositive question, while bracketing any inquiry into how nH Predict outputs in fact constrained that physician sign-off across the class. By fixing the procedural posture around UnitedHealth's preferred characterization of nH Predict as advisory (code 3), the motion sought to make the algorithm's role assessable without ever examining its operation across the class --- without override rates, denial patterns, or the structural performance-management constraints alleged in the complaint, by which UnitedHealth ``intentionally limit[ed] their employees' discretion to deviate from the nH Predict AI Model predication by setting up targets to keep stays at skilled nursing facilities within 1\% of the days projected by the AI Model'' and ``disciplined and terminated'' employees who deviated.\footnote{Class Action Compl.~\textparagraph\textparagraph~7, 35, \emph{The Estate of Gene B. Lokken v. UnitedHealth Grp., Inc.}, No. 23-cv-3514 (D. Minn. Nov. 14, 2023), ECF No. 1 [hereinafter \emph{Lokken} Compl.].} Layered over these request-side moves is a longstanding production-side posture invoking trade secrecy at the individual coverage-determination level (code 4): the complaint alleges that when patients and treating physicians request their nH Predict outcome reports, UnitedHealth's employees ``deny their requests and tell them that the information is proprietary.''\footnote{\emph{Lokken} Compl.~\textparagraph~37.}

Magistrate Judge Elkins denied the bifurcation motion on three grounds. First, the proposal assumed UnitedHealth's success on summary judgment; if its motion failed, plaintiffs would be made to litigate discovery and summary judgment twice.\footnote{\emph{The Estate of Gene B. Lokken}, 2025 WL 2607196, at *2.} Second, the court observed that UnitedHealth had resisted discovery throughout the litigation --- at the Rule 26(f) Report stage, at the Rule 16 scheduling conference, and through a successful motion to stay in August 2024 --- and was now, only after a partial adverse ruling on the motion to dismiss, moving to bifurcate for the first time, conduct that did not support a finding of good cause.\footnote{\emph{Id.}} Third, and most relevant, the court rejected the premise that this case turned on a ``narrow legal issue that might resolve a lurking dispositive issue,'' observing instead that UnitedHealth's proposal would ``litigate the entire case as to the named plaintiffs'' and that bifurcating would ``create more problems than it would solve.''\footnote{\emph{Id.} at *3.} The court noted that many of the discovery requests in plaintiffs' contemporaneously filed motion to compel were relevant to both the merits and class certification. 

\subsubsection{Mehrara v. Canada and Jahanian v. Canada}
\label{sec:chinook}
Chinook is a Microsoft Excel-based tool developed by IRCC in 2018 to process temporary resident applications at scale.\footnote{\emph{Mehrara v. Canada (Citizenship and Immigration)}, 2024 FC 1554 at para 8 (Can.).} It extracts applicant data from IRCC's Global Case Management System and presents it in a spreadsheet interface, allowing officers to review multiple applications simultaneously. Officers use ``Module 3'' to view information and ``Module 4'' to make the decision, the latter providing pre-generated reasons for refusal that can be tailored to an applicant's circumstances.\footnote{\emph{Id.} at para 9.} Chinook also surfaces risk indicators generated by a separate analytics tool called ITAT (Integrity Trends Analysis Tool). IRCC's official position is that Chinook is ``in essence an information management system'' that does not ``process, assess evidence, or make decisions on applications.''\footnote{\emph{Id.} at para 55.} \emph{Mehrara v. Canada}\footnote{\emph{Mehrara}, 2024 FC 1554.} and \emph{Jahanian v. Canada}\footnote{\emph{Jahanian}, 2024 FC 581.} are two 2024 Federal Court decisions arising from Chinook-assisted visa refusals; they are representative of a broader pattern of challenges beginning in 2023 in which applicants have repeatedly struggled to obtain access to the informational environment shaping their adjudications.

In \emph{Mehrara}, the applicants challenged IRCC's refusal of a study permit and associated temporary resident visa, arguing that Chinook 3+'s aggregation and presentation of their application data constrained the officer's independent judgment in undisclosed and unreviewable ways. The government's characterization of Chinook as a non-decision-making visual representation tool proved dispositive at every level of the access question (code~3): because the court accepted that framing, the informational environment Chinook created was rendered legally immaterial, and no production was required. Working notes were ruled ``only marginally relevant, if relevant at all,'' the ITAT risk score irrelevant because none had been assigned to the applicants, and the spreadsheets unnecessary because they merely organized information already in the GCMS.\footnote{\emph{Mehrara}, 2024 FC 1554 at paras 58, 60, 66.} The access determination was further foreclosed by IRCC's routine daily deletion of all Chinook-generated materials, meaning the spreadsheets no longer existed in any event (code~8). The court noted that this ``systematic daily deletion of all material generated by processing technology may not reflect best practice'' given that public power ``would be hollow without a basic understanding of how the exercise of that power occurs,'' but declined to order any change.\footnote{\emph{Id.} at para 69.} The government's characterization of Chinook as a mere organizational aid was never independently verified through production of the system itself (code~1).

In \emph{Jahanian}, the applicants raised the same Chinook arguments and additionally argued that Chinook's pre-set refusal language may channel what are in substance credibility findings --- which require the officer to contact the applicant --- into sufficiency-of-evidence refusals, which do not, bypassing procedural protections through interface design. Counsel conceded the Chinook arguments at the hearing as a new issue not properly raised at the leave stage.\footnote{\emph{Jahanian}, 2024 FC 581 at para 4.} No Chinook materials were produced or ordered, and the interface question was never adjudicated on the merits.

\section{Analysis of TrueAllele cases under different admissibility standards}
\label{sec:trueallele_all}
TrueAllele has been the subject of admissibility hearings in over fifty criminal proceedings across state and federal courts in the United States and internationally. Of these, we identified 24 cases in which source code access was actively contested. These cases span a range of evidentiary standards --- including Frye, Daubert, and state-specific standards such as Georgia's Harper standard and Virginia's Spencer hybrid standard --- providing an unusual opportunity to assess whether the admissibility standard applied to novel scientific evidence affects courts' willingness to compel disclosure of proprietary algorithmic evidence.

One might expect the choice of evidentiary standard to shape courts' willingness to compel source code access. Frye's general acceptance inquiry is arguably satisfiable through validation studies and expert testimony without requiring implementation-level access, while Daubert's explicit consideration of testability and error rates would seem to create stronger grounds for compelling disclosure. However, we find that courts have overwhelmingly declined to compel disclosure under both standards, suggesting that the privatization of proof dynamic operates largely independent of the admissibility standard applied to novel scientific evidence in criminal proceedings.

Table~\ref{tab:trueallele} summarizes outcomes across the 24 cases with active source code disputes. Courts declined to compel disclosure in 21 of 24 cases, and the pattern holds across all standards represented in the survey. \emph{State v. Simmer} illustrates the point most directly: the Nebraska Supreme Court under Daubert explicitly declined to require independent testing of the source code to support a general reliability finding, despite testability being a named Daubert factor and a central argument raised by the criminal defendant's counsel.\footnote{\emph{State v. Simmer}, 304 Neb. 369, 387 (2019).} Only three courts across all standards granted source code access: \emph{Pickett} (Frye, New Jersey), \emph{Ellis} (Daubert, federal), and \emph{Watson} (Spencer hybrid, Virginia). All three are among the more recent cases in this line of TrueAllele cases, and each involved specific arguments about how independent adversarial review of comparable software had revealed implementation errors not discoverable through other means, suggesting that if any trend exists it may be temporal rather than standard-dependent. This suggests that practical barriers to algorithmic evidence access, including trade secrecy, the sufficiency of developer-generated validation, and the absence of a clear procedural mechanism for compelling third-party disclosure, operate similarly regardless of the admissibility standard applied to novel scientific evidence in criminal proceedings.

\begin{table}[htbp]
    \centering
    {\color{black}
    \begin{tabular}{llcc}
        \toprule
        \textbf{Standard} & \textbf{Jurisdiction} & \textbf{Not compelled} & \textbf{Compelled} \\
        \midrule
        Frye                & State   & 12 & 1 \\
        Harper (Frye-like)  & GA      & 4  & 0 \\
        Spencer (hybrid)    & VA      & 2  & 1 \\
        Daubert             & State   & 3  & 0 \\
        Daubert             & Federal & 0  & 1 \\
        \midrule
        \textbf{Total}      &         & \textbf{21} & \textbf{3} \\
        \bottomrule
    \end{tabular}
    }
    \caption{TrueAllele source code disclosure outcomes across admissibility standards for scientific evidence}
    \label{tab:trueallele}
\end{table}

\subsection{Second example}
\label{sec:second_example}

Suppose an employer uses an automated screening system to rank or eliminate job applicants, and an applicant later sues after being rejected despite facially meeting the stated qualifications. She also alleges that the system screens out applicants from her protected group at disproportionately high rates.
The employer responds that the system is doing nothing more than sorting candidates according to legitimate business considerations and that any observed disparity is a byproduct of the applicant pool rather than unlawful discrimination. 
Step 1 would determine that there is significant asymmetry because the applicant has the adverse outcome and some initial evidence of disparity, but the employer controls the relevant data, system, and records needed to prove their claim. 
Thus, any request that is relevant would survive Step 1 because the applicant already has a cognizable discrimination claim but cannot fully rebut the employer's alternative explanation without additional access.

At Step 2, the test compares the applicant's requested access against the cause-of-action baseline---that is, the minimum additional access needed to reproduce the discrimination claim in a meaningful way. 
For example, ``minimal access needed to reproduce elements of the cause of action'' corresponds to what minimal information and access to the system would, together, reproduce the disproportionate rates the applicant observes.
This may be access sufficient to evaluate whether the challenged system treats similarly situated applicants differently, or whether it generated materially different screening results across protected groups during the relevant period. 
 In this case, query access to the system, information on how the employer uses the system (e.g., the input prompt), and possibly an anonymized sample of applications may be needed.
The plaintiff may prefer broader discovery, but broader requests should be assessed by comparing their added evidentiary value against their added production-side cost or risk. For instance, a request for complete source code, model weights, or the employer's full personnel archive will likely fail Step 2 because their incremental benefit may be lower relative to the production-side risk. 

In Step 3, the employer can offer substitutes. Imagine the plaintiff seeks direct interaction with the system so that she can run paired applications and observe whether slight  variations affect ranking or rejection. The employer may answer that the relevant version has been replaced, that live access would expose the system to manipulation, or that the tool is vendor-controlled and cannot simply be opened for adversarial testing. In response, the employer might propose alternatives such as a neutral expert review, a secure sandbox, preserved test results, or a controlled analysis of historical applicant-level data. These substitutes are acceptable only if they preserve the plaintiff's ability to surface supporting evidence, if it exists.

\end{document}